\documentclass[letterpaper,twocolumn,10pt]{article}
\usepackage{usenix}
\usepackage[noadjust]{cite}

  \usepackage[pdftex]{graphicx}
  \graphicspath{{images/}{images}}
  \DeclareGraphicsExtensions{.pdf,.jpeg,.png}

\usepackage{floatrow}
\usepackage{appendix}

\usepackage{amsmath}
\usepackage{textcomp}

\usepackage{multirow}
\usepackage{array}
\usepackage{booktabs}       %

\usepackage{stfloats}

\usepackage{caption}
\renewcommand{\thetable}{\Roman{table}}
\usepackage{url}
\usepackage{placeins}
 \makeatletter
 \DeclareRobustCommand\onedot{\futurelet\@let@token\@onedot}
 \def\@onedot{\ifx\@let@token.\else.\null\fi\xspace}

 \makeatother

\DeclareRobustCommand{\figref}[1]{Figure~\ref{#1}}

\DeclareRobustCommand{\Figref}[1]{Figure~\ref{#1}}

\DeclareRobustCommand{\secref}[1]{Section~\ref{#1}}

\DeclareRobustCommand{\Tableref}[1]{Table~\ref{#1}}
\DeclareRobustCommand{\tableref}[1]{Table~\ref{#1}}

\usepackage{xcolor}
\usepackage{adjustbox}
\usepackage{paralist}
\usepackage{microtype}

\newcommand{\authSpace}{\quad}

\newcommand{\myparagraph}[1]{\vspace{2pt}\noindent{\textbf{#1}}}

\newcolumntype{x}[1]{>{\raggedright\arraybackslash\hspace{0pt}}p{#1}}

\usepackage{lipsum}

\newcommand{\mario}[1]{}%
\newcommand{\bernt}[1]{}%
\newcommand{\raks}[1]{} %
\newcommand{\ul}[1]{\underline{#1}}
\newcommand{\ant}{{$\text{A}^{4}\text{NT }$}}
\newcommand{\antnospace}{{$\text{A}^{4}\text{NT}$}}
\newcommand{\ntext}[1]{{#1}}
\begin{document}
\title{
\antnospace: Author Attribute Anonymity\\
by Adversarial Training of Neural Machine Translation}

 \author{{\normalfont Rakshith Shetty \authSpace Bernt Schiele \authSpace Mario Fritz}\\
 Max Planck Institute for Informatics\\
 Saarland Informatics Campus\\
 Saarbr{\"u}cken, Germany\\
 {\normalfont Email: \url{firstname.lastname@mpi-inf.mpg.de}}}

\maketitle

\begin{abstract}
Text-based analysis methods enable an adversary to reveal privacy relevant author attributes such as gender, age and can identify the text's author. 
Such methods can compromise the privacy of an anonymous author even when the author tries to remove privacy sensitive content.
In this paper, we propose an automatic method, called Adversarial Author Attribute Anonymity Neural Translation (\antnospace), to combat such text-based adversaries.
\ntext{Unlike prior works on obfuscation, we propose a system that is fully automatic and learns to perform obfuscation entirely from data. This allows us to easily apply the \ant system to obfuscate different author attributes.}
We propose a sequence-to-sequence language model, inspired by machine translation, and an adversarial training framework to design a system which learns to transform the input text to obfuscate author attributes without paired data.
We also propose and evaluate techniques to impose constraints on our \ant 
to preserve the semantics of the input text.
\ant learns to make minimal changes to the input 
to successfully fool author attribute classifiers, while preserving the meaning of the input text. Our experiments on two datasets and three settings show that the proposed method is
effective in fooling the author attribute classifiers and thus 
improves the anonymity of authors.

\end{abstract}

\section{Introduction}

\ntext{Natural language processing (NLP) methods including stylometric tools
enable identification of authors of anonymous texts by analyzing stylistic properties of the text~\cite{juola2008authorship, stamatatos2009survey,ruder2016character}. 
NLP-based tools have also been applied to profiling users by determining their private attributes like age and gender~\cite{argamon2009automatically}. These methods have been shown to be effective in various settings like blogs, reddit comments, twitter text~\cite{overdorf2016blogs} and in large scale settings with up to 100,000 possible authors~\cite{narayanan2012feasibility}.} 
\ntext{In a recent famous case, authorship attribution tools were used to help confirm J.K Rowling as the real author of \emph{A Cuckoo's Calling} which was written by Ms. Rowling under pseudonymity~\cite{jkrowling}. This case highlights the privacy risks posed by these tools.}

\ntext{Apart from threat of identification of an anonymous author, the NLP-based tools also make authors susceptible to profiling. Text analysis has been shown to be effective in predicting age group~\cite{morgan2017predicting}, gender~\cite{Ikeda:2013} and to an extent even political preferences~\cite{makazhanov2014predicting}. By determining such private attributes an adversary can build user profiles which have been used for manipulation through targeted advertising, both for commercial and political goals~\cite{likesHelpedTrump}.}

\ntext{Since the NLP based profiling methods utilize the stylistic properties of the text to break the authors anonymity, they are immune to defense measures like pseudonymity, masking the IP addresses or obfuscating the posting patterns.}
\ntext{The only way to combat them is to modify the content of the text to hide stylistic attributes. Prior work has shown that while people are capable of altering their writing styles to hide their identity~\cite{brennan2012adversarial}, success rate depends on the authors skill and doing so consistently is hard for even skilled authors~\cite{afroz2012detecting}.}
Currently available solutions to obfuscate authorship and defend against NLP-methods has been largely restricted to semi-automatic solutions suggesting possible changes to the user~\cite{McDonaldACSG12} or hand-crafted transformations to text~\cite{castro:2017} which need re-engineering on different datasets~\cite{castro:2017}. This however limits the applicability of these defensive measures beyond the specific dataset it was designed on. To the best of our knowledge, text rephrasing using \ntext{generic machine translation} tools~\cite{keswani2016author} is the only prior work offering a fully automatic solution to author obfuscation which can be applied across datasets. But as \ntext{found in prior work~\cite{CaliskanTranslate2012} and further demonstrated} with our experiments, \ntext{generic machine translation} based obfuscation fails to sufficiently hide the identity and protect against attribute classifiers.

\ntext{Additionally the focus in  prior research has been towards protecting author identity. However, obfuscating identity does not guarantee protection of private attributes like age and gender. Determining attributes is generally easier than predicting the exact identity for NLP-based adversaries, mainly due to former being small closed-set prediction task compared to later which is larger and potentially open-set prediction task. This makes obfuscating attributes a difficult but an important problem.}

\myparagraph{Our work.} We propose an unified automatic system (\antnospace) to obfuscate authors text and defend against NLP adversaries. 
\ant follows the imitation model of defense discussed in~\cite{brennan2012adversarial} and protects against various attribute classifiers by learning to imitate the writing style of a target class. For example, \ant learns to hide the gender of a female author by re-synthesizing the text in the style of the male class. This imitation of writing style is learnt by adversarially training~\cite{goodfellow2014generative} our style-transfer network against the attribute classifier. Our \ant network learns the target style by learning to fool the authorship classifiers into mis-classifying the text it generates as target class. This style transfer is accomplished while aiming to retain the semantic content of the input text.

Unlike many prior works on authorship obfuscation~\cite{McDonaldACSG12, castro:2017}, we
propose an end-to-end learnable author anonymization solution, allowing us to apply our method not only to authorship obfuscation but to the anonymization of different author attributes including identity, gender and age with a {\it unified approach}.
We illustrate this by successfully applying our model on three different attribute anonymization settings on two different datasets. Through empirical evaluation, we show that the proposed approach is able to fool the author attribute classifiers in all three settings effectively and better than the baselines. \ntext{While there are still challenges to overcome before applying the system to multiple attributes and situations with very little data, we believe that \ant offers a new data driven approach to authorship obfuscation which can easily adapt to improving NLP-based adversaries.}

\myparagraph{Technical challenges:} We design our \ant network architecture based on the sequence-to-sequence neural machine translation model~\cite{seqToseq2014}. A key challenge in learning to perform style transfer, compared to other sequence-to-sequence mapping tasks like machine translation, is the lack of paired training data. Here, paired data refers to datasets with both the input text and its corresponding ground-truth output text. \ntext{In obfuscation setting, this means having a large dataset with semantically same sentences written in different styles corresponding to attributes we want to hide. Such paired data is infeasible to obtain and this has been a key hurdle in developing automatic obfuscation methods.}
Some prior attempts to perform text style transfer required paired training data~\cite{xu2012paraphrasing} and hence were limited in their applicability beyond toy-data settings.
We overcome this by training our \ant network within a generative adversarial networks (GAN)~\cite{goodfellow2014generative} framework. GAN framework enables us to train the \ant network to generate samples that match the target distribution without need for paired data.

We characterize the performance of our \ant network along two axes: privacy effectiveness and
semantic similarity. Using automatic metrics and human evaluation to measure semantic similarity of the generated text to the input, we show that \ant offers a better trade-off between privacy effectiveness and
semantic similarity. We also analyze the effectiveness of \ant for protecting anonymity for varying degrees of input text ``difficulty''.

\myparagraph{Contributions:}
In summary, the main contributions of our paper are. %
{\bf (1):} %
We propose a novel approach to authorship obfuscation, that uses a style-transfer network (\antnospace) to automatically transform the input text to a target style and fool the attribute classifiers. The network is trained without paired data using adversarial training.
{\bf(2):} %
The proposed  obfuscation solution is end-to-end trainable, and hence can be applied to protect different author attributes and on different datasets with no changes to the overall framework.
{\bf (3):} %
Quantifying the performance of our system on privacy effectiveness and semantic similarity to input, we show that it offers a better trade-off between the two metrics compared to baselines. %

\section{Related Work} 
In this section, we review prior work relating to four different aspects of our work -- author attribute detection~(our adversaries), authorship obfuscation~(prior work), machine translation~(basis of our \ant network) and 
generative adversarial networks~(training framework we use).

\myparagraph{Authorship and attribute detection}
Machine learning approaches where a set of text features are input to a classifier which learns to predict the author have been popular in recent author attribution works~\cite{stamatatos2009survey}. These methods have been shown to work well on large datasets~\cite{narayanan2012feasibility}, duplicate author detection ~\cite{afroz2014doppelganger} and even on non-textual data like code~\cite{caliskan2015anonymizing}.  Sytlometric models can also be applied to determine private author attributes like age or gender~\cite{argamon2009automatically}. %

Classical author attribution methods rely on a predefined set of features extracted from the input text~\cite{abbasi2008writeprints}.
Recently deep-learning methods have been applied to learn to extract the features directly from  data~\cite{bagnall2015author,ruder2016character}. \cite{bagnall2015author} uses a multi-headed recurrent neural network~(RNN) to train a generative language model on each author's text and use the model's perplexity on the test document to predict the author. Alternatively, \cite{ruder2016character} uses convolutional neural network~(CNN) to train an author classifiers. To show generality of our \ant network, we test it against both RNN and CNN based author attribute classifiers.

\myparagraph{Authorship obfuscation}
Authorship obfuscation methods are adversarial in nature to stylometric methods of author attribution; they try to change the style of input text so that author identity is not discernible.
The majority of prior works on author attribution are semi-automatic~\cite{obfGary2006, McDonaldACSG12}, where the system suggests authors to make changes to the document by analyzing the stylometric features.
The few automatic obfuscation methods have relied on general rephrasing methods like \ntext{generic} machine translation~\cite{keswani2016author} or on a predefined text transformations~\cite{karadzhov2017case}. Round-trip machine translation, where input text is translated to multiple languages one after the other until it is translated back to the source language, is proposed as an automatic method of obfuscation in \cite{keswani2016author}. Recent work \cite{karadzhov2017case} obfuscates text by moving the stylometric features towards the average values on the dataset applying pre-defined transformations on input text.

\ntext{We propose the first method to achieve fully automatic obfuscation using text style transfer. This style transfer is not pre-defined but learnt directly from data optimized for fooling attribute classifiers. This allows us to apply our model across datasets without extra engineering effort.}

\myparagraph{Machine translation} The task of style-transfer of text data shares similarities with the machine translation problem. Both involve mapping an input text sequence onto an output text sequence. 
Style transfer can be thought of as machine translation on the same language.

Large end-to-end trainable neural networks %
have become a popular choice in machine translation~\cite{bahdanau2014neural,wu2016google}. 
These methods are generally based on sequence-to-sequence recurrent models~\cite{seqToseq2014} consisting of two networks, an encoder which encodes the input sentence into a fixed size vector and a decoder which maps this encoding to a sentence in the target language.
We base our \ant network architecture on the word-level sequence-to-sequence language model~\cite{seqToseq2014}. Neural machine translation systems are trained with large amounts of paired training data. %
However, in our setting, obtaining paired data of the same text in different writing styles is not viable.
We overcome the lack of paired data by casting the task as matching style distributions instead of matching individual sentences. Specifically, our \ant network takes an input text from a source distribution and generates text whose style matches the target attribute distribution. This is learnt without paired data using distribution matching methods.
\ntext{This reformulation allows us to demonstrate the first successful application of the machine translation models to the obfuscation task.}

\myparagraph{Generative adversarial networks} Generative Adversarial
Networks~(GAN)~\cite{goodfellow2014generative} are a framework for learning a generative model to produce samples from a target distribution. It consists of two models, a generator and a discriminator. %
The discriminator network learns to distinguish between the generated samples and real data samples. Simultaneously, the generator learns to fool this discriminator network thereby getting closer to the target distribution. In this two-player game, a fully optimized generator perfectly mimics the target distribution~\cite{goodfellow2014generative}.

We train our \ant network within the GAN framework, directly optimizing \ant to fool the attribute classifiers by matching style distribution of a target class. A recent approach to text style-transfer proposed in~\cite{shen2017style} also utilizes GANs to perform style transfer using unpaired data. However, the solution proposed in~\cite{shen2017style} changes the meaning of the input text significantly during style transfer and is applied on sentiment transfer task.
In contrast, authorship obfuscation task requires the generated text to preserve the semantics of the input.
We address this problem by proposing two methods to improve semantic consistency between the input and the \ant output.
\section{Author Attribute Anonymization}

\begin{figure}
\centering
    \includegraphics[width=0.80\columnwidth]{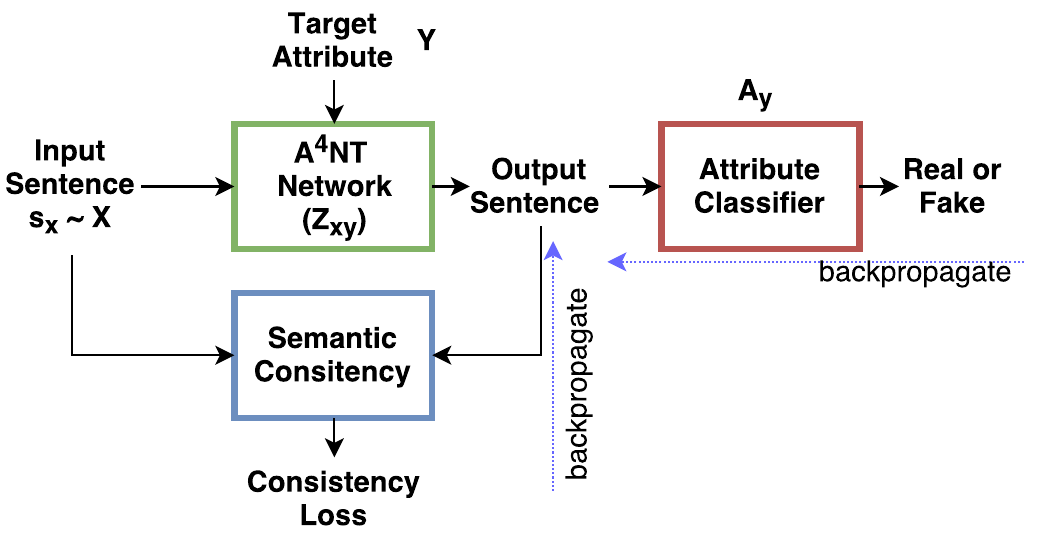}
    \caption{GAN framework to train our \ant network. Input sentence is transformed by \ant to match the style of the target attribute. This output is evaluated using the attribute classifier and semantic consistency loss. \ant is trained by backpropagating through these losses.\vspace{-4mm}}
\label{fig:OverallModel}
\end{figure}
We propose an author adversarial attribute anonymizing neural translation~(\antnospace) network to defend against NLP-based adversaries.
The proposed solution includes the \ant Network , the adversarial training scheme, and semantic and language losses to learn to protect private attributes.
The \ant network transforms the input text from a source attribute class to mimic the style of a different attribute class, and thus fools the attribute classifiers.

Technically, \ant network is essentially solving a sequence to sequence mapping problem --- from text sequence in the source domain to text in the target domain --- similar to machine translation.
Exploiting this similarity, we design our \ant network based on the sequence-to-sequence neural language
models~\cite{seqToseq2014}, widely used in neural machine translation~\cite{bahdanau2014neural}.
These models have proven effective when trained with large amounts of paired data and are also deployed  commercially~\cite{wu2016google}.
If there were paired data in source and target attributes, we could train our \ant network exactly like a machine translation model, with standard supervised learning. 
However, such paired data is infeasible to obtain as it would require the same text written in multiple styles.

To address the lack of paired data, we cast the anonymization task as learning a generative model, $Z_{xy}(s_x)$, which
transforms an input text sample $s_x$ drawn from source attribute distribution $s_x\sim X$, to look like samples from the target distribution $s_y\sim Y$.
This formulation enables us to train the \ant network $Z_{xy}(s_x)$ with the GAN framework to produce samples close to the target distribution $Y$, %
using only unpaired samples from $X$ and $Y$. \Figref{fig:OverallModel} shows this overall framework.%
The GAN framework consists of two models, a generator producing synthetic samples to mimic the target data distribution, and a discriminator which tries to distinguish
real data from the synthesized ``fake'' samples from the generator.
The two models are trained adversarially%
, i.e. the generator tries to fool the discriminator and the discriminator tries to correctly
identify the generator samples.
We use an attribute classifier as the discriminator and the \ant network as the
generator.
The \ant network, in trying to fool the attribute classification network, learns to transform the input text to mimic the style of the target attribute and protect the attribute anonymity. 

For our \ant network to be a practically useful defensive measure, the text output by this network should be able to fool the attribute classifier while also preserving the meaning of the input sentence.
If we could measure the semantic difference between the generated text and the input text it could be used to penalize deviations from the input sentence semantics. 
Computing this semantic distance perfectly would need true understanding of the meaning of input sentence, which is beyond the capabilities of current natural language processing techniques.
To address this aspect of style transfer, we experiment with various proxies to measure and penalize changes to input semantics,
which will be discussed in \secref{sec:semanticloss}.
Following subsections will describe each module in detail. 

\subsection{Author Attribute Classifiers}
\label{sec:auhtclassifier}
\begin{figure}
\centering
    \includegraphics[width=0.95\columnwidth]{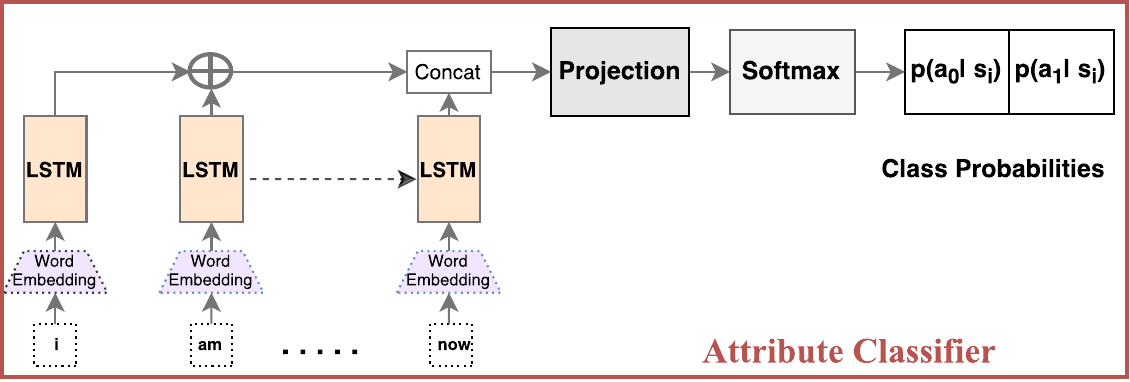}
    \caption{Block diagram of the attribute classifier network. The LSTM encoder embeds the input sentence into a vector. Sentence encoding is passed to linear projection followed by
softmax layer to obtain class probabilities\vspace{-3mm}}
\label{fig:EvalModel}
\end{figure}
We build our attribute classifiers using neural networks that predict the attribute label by directly operating on the text data. This is similar to recent approaches in authorship recognition~\cite{bagnall2015author,ruder2016character} where, instead of hand-crafted features used in classical stylometry, neural networks are used to directly predict author identity from raw text data. However, unlike in these prior works, our focus is attribute classification and obfuscation. We train our classifiers with recurrent networks operating at word-level, as opposed to character-level models used in~\cite{bagnall2015author,ruder2016character} for two reasons. We found that the word-level models give good performance on all three attribute-classification tasks we experiment with (see \secref{sec:quantresults}). Additionally, they are much faster than character-level models, making it feasible to use them in GAN training described in~\secref{sec:GeneratorModel}.  

Specifically, our attribute classifier $A_x$ to detect attribute value $x$ is shown in \Figref{fig:EvalModel}. 
It consists of a Long-Short Term Memory~(LSTM)~\cite{hochreiter1997long} encoder network to compute an embedding of the input sentence into a fixed size vector. 
It learns to encode the parts of the sentence most relevant to the classification task into the embedding vector, which for attribute prediction is mainly the stylistic properties of the text. This embedding is input to a linear layer and a softmax layer to output the class probabilities.

Given an input sentence $s_x=\{w_0, w_1, \cdots, w_{n-1}\}$, the words are one-hot encoded and then embedded into fixed size vectors using the word-embedding layer shown in~\Figref{fig:EvalModel} to obtain vectors $\{v_0, v_1,\cdots, v_{n-1}\}$. This word embedding layer encodes similarities between words into the word vectors and can help deal with large vocabulary sizes. The word vectors are randomly initialized and then learned from the data during training of the model. This approach works better than using pre-trained word vectors like word2vec~\cite{mikolov13nips} or Glove~\cite{pennington2014glove} since the learned word-vectors can encode similarities most relevant to the attribute classification task at hand.

This sequence of word vectors is recursively passed through an LSTM to obtain a sequence of outputs $\{h_0, h_1, \cdots, h_{n-1}\}$. We refer the reader to~\cite{hochreiter1997long} for the exact computations performed to get the LSTM output.

Now sentence embedding is obtained by concatenation of the final LSTM output and the mean of the LSTM outputs from other time-steps.
\begin{align}
  \label{eqn:sentenc}
  E(s_x) = \left\lbrack h_{n-1} ; \frac{1}{n-1}\sum{h_{n-1}}\right\rbrack
\end{align}
At the last time-step the LSTM network has seen all the words in the sentence and can encode a summary of the sentence in its output. However, using LSTM outputs from all time-steps, instead of just the final one, speeds up training due to improved flow of gradients through the network. 
Finally, $E(s_x)$ is passed through linear and softmax layers to obtain class probabilities, for each class $c_i$. The network is then trained using cross-entropy loss.
\begin{align}
  \label{eqn:discrmLoss}
  p_{\text{auth}}(c_i|s_x) &= \text{softmax}(W\cdot E(s_x))\\
  \text{Loss}(A_x) &=  \sum_{i}{t_i(s_x)\log{\left(p_{\text{auth}}(c_i|s_x)\right)}}
\end{align}
\noindent
where $t(s_x)$ is the one-hot encoding of the true class of $s_x$.%

The same network architecture is applied for all our attribute prediction tasks including identity, age and gender.%

\subsection{The \ant Network}
\label{sec:GeneratorModel}
\begin{figure}
\centering
    \includegraphics[width=0.97\columnwidth]{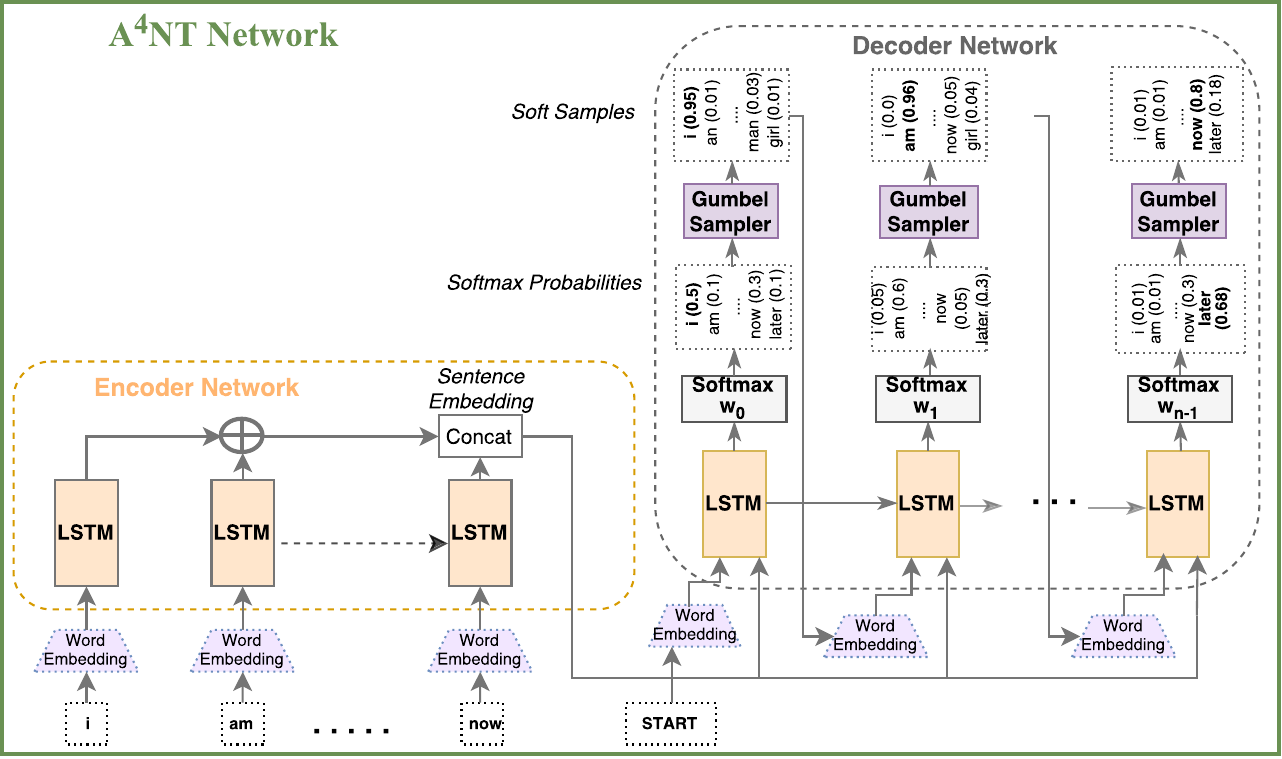}
    \caption{Block diagram of the \ant network. %
    First LSTM encoder embeds the input sentence into a vector. The decoder maps this sentence encoding to the output sequence. Gumbel sampler produces ``soft'' samples from the softmax distribution to allow backpropagation.\vspace{-5mm}}
\label{fig:GeneratorModel}
\end{figure}
A key design goal for the \ant network is that it is trainable purely from data to obfuscate the author attributes.%
This is a significant departure from prior works on author obfuscation~\cite{McDonaldACSG12, karadzhov2017case} that rely on hand-crafted rules for text modification to achieve obfuscation. The methods relying on hand-crafted rules are limited in applicability to specific datasets they were designed for.%

To achieve this goal, we base our \ant network $Z_{xy}$, shown in \Figref{fig:GeneratorModel}, on a recurrent sequence-to-sequence neural translation model~\cite{seqToseq2014}~(\emph{Seq2Seq}) popular in many sequence mapping tasks. 
As seen from the wide-range of applications mapping text-to-text~\cite{bahdanau2014neural}, speech-to-text~\cite{weiss2017sequence}, text-to-part of speech~\cite{ma2016end}, the \emph{Seq2Seq} models can effectively learn to map input sequences to arbitrary output sequences, with appropriate training. They operate on raw text data and alleviate the need for hand-crafted features or rules to transform the style of input text, predominantly used in prior works on author obfuscation~\cite{McDonaldACSG12, karadzhov2017case}. Instead, appropriate text transformations can be learnt directly from  data. 
This flexibility allows us to easily apply the same \ant network and training scheme to different datasets and settings. 

The \ant network $Z_{xy}$ consists of two components, an encoder and a decoder modules, similar to standard sequence-to-sequence models.
The encoder embeds the variable length input sentence into a fixed size vector space. The decoder maps the vectors in this embedding space to output text sequences in the target style.
The encoder is an LSTM network, sharing the architecture of the sentence encoder in \secref{sec:auhtclassifier}. The same architecture applies here as the task here is also to embed the input sentence $s_x$ into a fixed size vector $E_{G}(s_x)$.
However, $E_{G}(s_x)$ should learn to represent the semantics of the input sentence allowing the decoder network to generate a sentence with similar meaning but in a different style.

The sentence embedding from the encoder is the input to the decoder LSTM %
which generates the output sentence one word at a time.
At each step $t$, the decoder LSTM takes $E_{G}(s_x)$ and the previous output word $w^{o}_{t-1}$ to produce a probability distribution over the vocabulary. Sampling from this distribution outputs the next word.
\begin{align}
  h^{\text{dec}}_t(s_x) &= \text{LSTM}\left\lbrack E_{G}(s_x),W_{\text{emb}}(\tilde{w}_{t-1})\right\rbrack\\
  \label{eqn:genOut}
  p(\tilde{w}_t|s_x) &= \text{softmax}_{V}(W_{\text{dec}}\cdot h^{\text{dec}}_t(s_x))\\
  \tilde{w}_t &= \text{sample}(p(\tilde{w}_t|s_x))
\end{align}
\noindent where $W_{\text{emb}}$ is the word embedding, $W_{\text{dec}}$ matrix maps the LSTM output to vocabulary size and $V$ is the vocabulary.

In most applications of \emph{Seq2Seq} models, the networks are trained using parallel training data, consisting of input and ground-truth output sentence pairs. 
A sentence is input to the encoder and propagated through the network and the network is trained to maximize the likelihood of generating the paired ground-truth output sentence. 
However, in our setting, we do not have access to such parallel training data of text in different styles and the \ant network $Z_{xy}$ is trained in an unsupervised setting.

We address the lack of parallel training data by using the GAN framework to train the \ant network. In this framework, the \ant network $Z_{xy}$ learns by generating text samples and improving itself iteratively to produce text that the attribute classifier, $A_y$, classifies as target attribute. A benefit of GANs is that the \ant network is directly optimized to fool the attribute classifiers. It can hence learn to make transformations to the parts of the text which are most revealing of the attribute at hand, and so hide the attribute with minimal changes.

However, to apply the GAN framework, we need to differentiate through the samples generated by $Z_{xy}$. 
The word samples from $p(\tilde{w}_t|s_x)$ are discrete tokens and are not differentiable. Following~\cite{Shetty2017iccv}, we apply the Gumbel-Softmax approximation~\cite{jang2016categorical} to obtain differentiable soft samples and enable end-to-end GAN training. See Appendix~\ref{sec:appendixA} for details.

\myparagraph{Splitting decoder:} To transfer styles between attribute pairs, $x$ and $y$, in both directions, we found it ineffective to use the same network $Z_{xy}$. A single network $Z_{xy}$ is unable to sufficiently switch its output word distributions solely on a binary condition of target attribute. Nonetheless, using a separate network for each ordered pair of attributes is prohibitively expensive. A good compromise we found is to share the encoder to embed the input sentence but use different decoders for style transfer between each ordered pair of attributes.
Sharing the encoder allows the two networks to share a significant number of parameters and enables the attribute specific decoders to deal with words found only in the vocabulary of the other attribute group using shared sentence and word embeddings.
\subsection{Style Loss with GAN}%
\begin{figure}
\centering
    \includegraphics[width=.75\columnwidth]{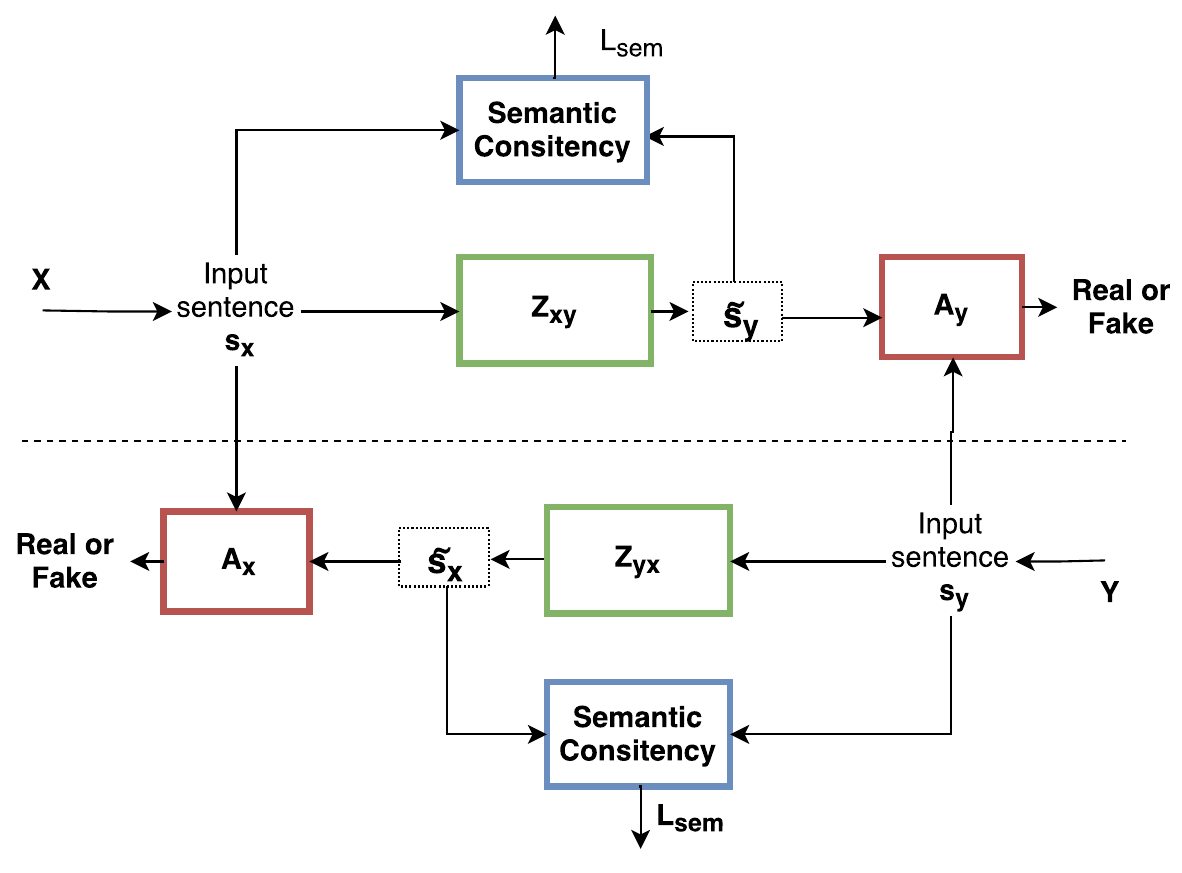}\vspace{-1mm}
    \caption{Illustrating use of GAN framework and cyclic semantic loss to train a pair of \ant networks.\vspace{-3mm}}%
\label{fig:GanTraining}
\end{figure}

We train the two \ant networks $Z_{xy}$ and $Z_{yx}$ in the GAN framework
to produce samples which are indistinguishable from samples from distributions of attributes $y$ and $x$ respectively, without having paired sentences from $x$ and $y$.
\Figref{fig:GanTraining} shows this training framework.%

Given a sentence $s_x$ written by author with attribute $x$, the \ant network outputs a sentence $\tilde{s_y}=Z_{xy}(s_x)$.
This is passed to the attribute classifier for attribute $y$, $A_y$, to obtain probability $p_{\text{auth}}(y|\tilde{s_y})$.
$Z_{xy}$ tries to fool the classifier $A_y$ into assigning high probability to its output,
whereas $A_y$ tries to assign low probability to sentences produced by $Z_{xy}$ while assigning high probability to real sentences $s_y$ written by $y$.
The same process is followed to train the \ant network from $y$ to $x$, with $x$ and $y$ swapped.
The loss functions used to train the \ant network and the attribute classifiers in this setting is given by:

\begin{align}
  \label{eqn:stylegan}
  &L(A_y) = -\log\left({p_{\text{auth}}(y|s_y)}\right) - \log\left({1 - p_{\text{auth}}(y|\tilde{s_y})}\right)\\
  &L_{\text{style}}(Z_{xy}) = - \log\left(p_{\text{auth}}(y|\tilde{s_y})\right)
\end{align}

The two networks $Z_{xy}$ and $A_y$ are adversarially competing with each other when minimizing the above loss functions. At optimality it is guaranteed that the distribution of samples produced by $Z_{xy}$ is identical to the distribution of $y$~\cite{goodfellow2014generative}. However, we want the \ant network to only imitate the style of $y$, while keeping the content from $x$. Thus, we explore methods to enforce the semantic consistency between the the input sentence and the \ant output.
\subsection{Preserving Semantics}
\label{sec:semanticloss}
We want the output sentence, $\tilde{s_y}$, produced by $Z_{xy}(s_x)$ to not only fool the attribute classifier, but also to preserve the meaning of the input sentence $s_x$.
We propose a semantic loss $L_{\text{sem}}(\tilde{s_y}, s_x)$ to quantify the meaning changed during the anonymization by \ant.
Simple approaches like matching words in $\tilde{s_y}$ and $s_x$ can severely limit the effectiveness of anonymization, as it penalizes even synonyms or alternate phrasing. 
In the following subsection we will discuss two approaches to define $L_{\text{sem}}$, and later
in \secref{sec:results} we compare these approaches quantitatively.

\subsubsection{Cycle Constraints}
\label{sec:cycsemloss}
\begin{figure}
\centering
    \includegraphics[width=0.8\columnwidth]{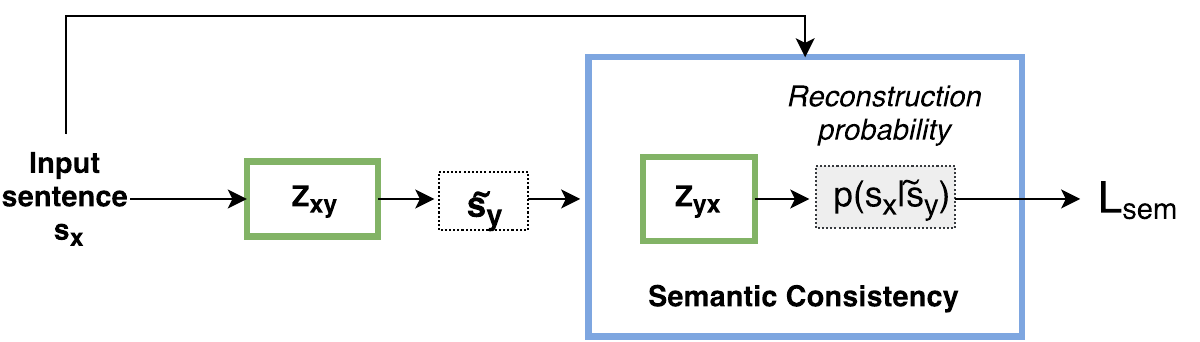}\vspace{-1mm}
    \caption{Semantic consistency in \ant networks is enforced by maximizing cyclic reconstruction probability.\vspace{-3mm}}%
\label{fig:SemConsist}
\end{figure}
One could evaluate how semantically close is $\tilde{s_y}$ to $s_x$ by evaluating how easy it is to
reconstruct $s_x$ from $\tilde{s_y}$.
If $\tilde{s_y}$ means exactly the same as $s_x$, there should be no information loss and we should be able to perfectly reconstruct $s_x$ from $\tilde{s_y}$.
We could use the \ant network in the reverse direction to obtain a reconstruction,
$\ddot{s_x}=Z_{yx}(\tilde{s_y})$ and compare it to input sentence $s_x$.
Such an approach, referred to as cycle constraint, has been used in image style transfer~\cite{CycleGAN2017}, where $l_1$ distance is used to compare the reconstructed image and the original image to impose semantic relatedness penalty. 
However, in our case $l_1$ distance is not meaningful to compare $\ddot{s_x}$ and $s_x$, as they are sequences of possibly different lengths. Even a single word insertion or deletion in $\ddot{s_x}$ can cause the entire sequence to mismatch and be penalized by the $l_1$ distance. 

A simpler and more stable alternative we use is to forgo the reconstruction and just
compute the likelihood of reconstruction of $s_x$ when applying reverse style-transfer on $\tilde{s_y}$. 
This likelihood is simple to obtain from the reverse \ant network $Z_{yx}$ using the word distribution probabilities at the output.
This cyclic loss computation is illustrated in \Figref{fig:SemConsist}.
Duly, we compute reconstruction probability $P_r(s_x |\tilde{s_y})$ and define the semantic loss as:
\begin{align}
  \label{eqn:reconstructionloss}
  P_r(s_x | \tilde{s_y}) &=  \prod_{t=0}^{n-1} p_{z_{yx}}(w_t|\tilde{s_y})\\
  \label{eqn:cycloss}
  L_{\text{sem}}(\tilde{s_y},s_x)&=-\log{P_r(s_x | \tilde{s_y})}
\end{align}
\noindent The lower the semantic loss $L_{\text{sem}}$, the higher the reconstruction probability and thus more meaning of the input sentence $s_x$ is preserved in the style-transfer output $\tilde{s_y}$. %

\subsubsection{Semantic Embedding Loss}
\label{sec:semembloss}
Alternative approach to measure the semantic loss is to embed the two sentences, $\tilde{s_y}$ and $s_x$, into a semantic space and compare the two embedding vectors using $l_1$ distance. %
The idea is that a semantic embedding method puts similar meaning sentences close to each other in this vector space.
This approach is used in many natural language processing tasks, for example in semantic entailment~\cite{conneau2017supervised}

Since we do not have annotations of semantic relatedness on our datasets, it is not possible to train a semantic embedding model but instead we have to rely on pre-trained models known to have good transfer learning performance. Several such semantic sentence embeddings are available in the
literature~\cite{kiros2015skip,conneau2017supervised}. 
We use the universal sentence embedding model from~\cite{conneau2017supervised}, pre-trained on
the Stanford natural language inference dataset~\cite{bowman2015large}.

We embed the two sentences using this semantic embedding model $F$ and use the $l_1$ distance to
compare the two embeddings and define the semantic loss as:
\begin{align}
  \label{eqn:encloss}
  L_{\text{sem}}(\tilde{s_y},s_x)&= \sum_{dim}{\left|F(s_x)-F(\tilde{s_y})\right|}
\end{align}

\subsection{Smoothness with Language Loss}
\label{sec:langloss}

The \ant network can minimize the style and the semantic losses, while still producing text which is broken and grammatically incorrect.
To minimize the style loss the \ant network needs to add words typical of the target attribute style, while  minimizing the semantic loss, it needs to retain the semantically relevant words from the input text.
However neither of these two losses explicitly enforces correct grammar and word order of $\tilde{s}$.

On the other hand, unconditional neural language models are good at producing grammatically correct text.
The likelihood of the sentence produced by our \ant model $\tilde{s}$ under an
unconditional language model, $M_y$, trained on the text by target attribute authors $y$, is a good indicator of the
grammatical correctness of $\tilde{s}$.
The higher the likelihood, the more likely the generated text $\tilde{s}$ has syntactic properties seen in the real data.
Therefore, we add an additional language smoothness loss on $\tilde{s}$ in order to enforce $Z$ to produce syntactically correct text.
\begin{align}
  \label{eqn:langloss}
  L_{\text{lang}}(\tilde{s})&= -\log{M_y(\tilde{s})} 
\end{align}

\myparagraph{Overall loss function:}
The \ant network is trained with a weighted combination of the three losses: style loss, semantic consistency and language smoothing loss.
\begin{align}
  \label{eqn:lossTotal}
  L_{\text{tot}}(Z_{xy})&= w_{\text{sty}}L_{\text{style}} + w_{\text{sem}}L_{\text{sem}} + w_{l}L_{\text{lang}}
\end{align}
\noindent We chose the above three weights so that the magnitude of the weighted loss terms are approximately equal at the beginning of training. Model training was not sensitive to exact values of the weights chosen that way.

\myparagraph{Implementation details:} We implement our model using the Pytorch framework~\cite{pytorch}. The networks are trained by optimizing the loss functions described with stochastic gradient descent using the RMSprop algorithm~\cite{tieleman2012lecture}. The \ant network is pre-trained as an autoencoder, i.e to reconstruct the input sentence, before being trained with the loss function described in~\eqref{eqn:lossTotal}. During GAN training, the \ant network and the attribute classifiers are trained for one minibatch each alternatively. We will open source our code, models and data at the time of publication.
\section{Experimental Setup}\label{sec:exp-setup}
We test our \ant network on obfuscation of three different  attributes of authors on two different datasets.
The three attributes we experiment with include author's age (under 20 vs over 20), gender (male vs female authors),  and author identities (setting with two authors).

\subsection{Datasets}
We use two real world datasets for our experiments: Blog Authorship corpus~\cite{schler2006effects} and Political Speech dataset. The datasets are from very different sources with distinct language styles, the first being from mini blogs written by several anonymous authors, and the second from political speeches of two US presidents Barack Obama and Donald Trump. This allows us to show that our approach works well across very different language corpora.

\myparagraph{Blog dataset:} The blog dataset is a large collection of micro blogs from  blogger.com collected by \cite{schler2006effects}. The dataset consists of 19,320 ``documents'' along with annotation of author's age, gender, occupation and star-sign. Each document is a collection of all posts by a single author. We utilize this dataset in two different settings; split by gender~(referred to as blog-gender setting) and split by age annotation~(blog-age setting). In the blog-age setting, we group the age annotations into two groups, teenagers~(age between 13-18) and adults~(age between 23-45) to obtain data with binary age labels. Age-groups 19-22 are missing in the original dataset. Since the dataset consists of free form text written while blogging with no proper sentence boundaries markers, we use the Stanford CoreNLP tool to segment the documents into sentences. All numbers are replaced with the NUM token. 

\myparagraph{Political speech dataset:} To test the limits of how far style imitation based anonymization can help protect author identity, we also test our model on two well known political figures with very different verbal styles. We collected the transcriptions of political speeches
of Barack Obama and Donald Trump made available by the The American Presidency Project~\cite{woolley2008american}. While the two authors talk about similar topics they have highly distinctive styles and vocabularies, making it a challenging dataset for our \ant network.
The dataset consists of 372 speeches, with about 65,000 sentences in total as shown in \Tableref{tab:DatasetComparison}. We treat each speech as a separate document when evaluating the classification results on document-level.
This dataset contains a significant amount of references to named entities like people, organizations, etc. To avoid that both attribute classifiers and the style transfer model rely on these references to specific people, we use the Stanford Named Entity Recognizer tool~\cite{finkel2005incorporating} to identify and replace these entities with entity labels.

The comparison of the two datasets can be found in \Tableref{tab:DatasetComparison}. The blog dataset is much larger and therefore we run most of our evaluation on it. Using these two datasets, we evaluate our model in three different attribute obfuscation settings, namely age~(blog-age), gender~(blog-gender) and identity obfuscation~(speech dataset). Detailed analysis of our model presented in \secref{sec:qualresults} is done on the validation split of the blog dataset, in the blog-age setting, containing 2,799 documents and 518,268 sentences. 

\begin{table}[t]
\footnotesize
\centering
\caption{Comparing statistics of the two datasets.\vspace*{-5mm}} 
\label{tab:DatasetComparison}
\hspace*{-5mm}\begin{tabular}{lcccc}
\toprule
    \bf Dataset& Attributes &\# Documents& \# Sentences &\# Vocabulary\\\cmidrule(lr){1-1}\cmidrule(lr){2-5}
Speech & Identity & 372 &  65k& 5.6k \\
Blog & Age, Gender & 19320 & 3.38 Mil  & 22k\\
 \bottomrule
\end{tabular}\vspace{-2mm}
\end{table}

\subsection{Evaluation Methods}

We evaluate our models w.r.t. two different aspects: \textbf{(a)} \emph{Privacy effectiveness}, to measure how well the \ant network can fool the attribute classifiers and \textbf{(b)} \emph{Semantic similarity}, quantifying how much of the input sentence semantics is preserved after style transfer.

To quantify privacy effectiveness we compare the F1-score of the attribute classifiers on the original data and the generated text. To mimic a real-world setting where we will not have access to the parameters of the attribute-classifier used by the adversary to train our \ant network, we test it against attribute classifiers which have been separately trained on real data.
The F1-score for the attribute classifiers is defined as:
\begin{align}
  \label{eqn:F1score}
  F1 &= 2*\frac{\text{precision} * \text{recall}}{\text{precision}+\text{recall}}
\end{align}
With successful style transfer, the attribute classifier will mis-classify generated text and the F1-scores will drop. We consider the obfuscation to be successful if the F1-score drops below that of a a random guessing classifier. Note that since the datasets are not perfectly balanced w.r.t the binary attributes, the F1-score for random classifier is not 0.5. The expected document-level F1-scores of random guessing are 0.54 for the blog-age setting, 0.5 for the blog-gender setting, and 0.61 for the speech data.

To quantify semantic similarity, we use the Meteor~\cite{denkowski-lavie:2014:Meteor} metric.
It is used in machine translation and image captioning to evaluate the similarity of the candidate text using a reference text.
Meteor compares the candidate text to one or more references by matching n-grams, allowing for soft matches using synonym and paraphrase tables.
We use the Meteor score between the generated and input text as the measure of semantic similarity.

However, the automatic evaluation for semantic similarity is not perfectly correlated with human judgments, especially with few reference sentences.%
To address this, we additionally conduct a human evaluation study on a subset of the test data of 745 sentences. We ask human annotators on Amazon Mechanical Turk to judge the semantic similarity of the generated text from our models. No other information was collected from the annotators, thereby keeping them anonymous. The annotators were compensated for their work through AMT system. We manually screened the text shown to the annotators to make sure there was no obvious offensive content in them.

\subsection{Baselines}
\label{sec:base}
We use the two baseline methods below to compare our model with. Both chosen baselines are automatic obfuscation methods not relying on hand-crafted rules.%

\myparagraph{Autoencoder} We train our \ant network $Z$ as an autoencoder, where it takes as input $s_x$ and tries to reproduce it from the encoding. The autoencoder is trained similar to a standard neural language model with cross entropy loss. We train two such auto-encoders $Z_{xx}$ and $Z_{yy}$ for the two attributes. Now simple style transfer can be achieved from $x$ to $y$ by feeding the sentence $s_x$ to the autoencoder of the other attribute class $Z_{yy}$. Since $Z_{yy}$ is trained to output text in the $y$ domain, the sentence $Z_{yy}(s_x)$ tends to look similar to sentences in $y$. This model sets the baseline for style transfer that can be achieved without cross domain training using GANs, with the same network architecture and the same number of parameters.

\myparagraph{Google machine translation:} A simple and accessible approach to change writing style of a piece of text without hand designed rules is to use generic machine translation software. The input text is translated from a source language to multiple intermediate languages and finally translating back to the source language. The hope is that through this round-trip the style of the text has changed, with the meaning preserved. This approach was used in the PAN authorship obfuscation challenge recently~\cite{keswani2016author}.%

We use the Google machine translation service\footnote{https://translate.google.com/} to perform round-trip translation on our input sentences. 
We have tried a varying number of intermediate languages, results of which will be discussed in~\secref{sec:results}
Since Google limits api-calls and imposes character limits on manual translation, we use this baseline only on the subset of 745 sentences from the test set for human evaluation.%

\section{Experimental Results}
\label{sec:results}

We test our model on the three settings discussed in section \ref{sec:exp-setup} with the goal to understand if the proposed \ant network can fool the attribute classifiers to protect the anonymity of the author attributes. 
Through quantitative evaluation done in \secref{sec:quantresults}, we show that this is indeed the case: our \ant network learns to fool the attribute classifiers across all three settings. We compare the two semantic loss functions presented in \secref{sec:semanticloss} and show that the proposed reconstruction likelihood loss does better than pre-trained semantic encoding.

However, this privacy gain comes with a trade-off. The semantics of the input text is sometimes altered. In \secref{sec:qualresults}, using qualitative examples, we analyze the failure modes of our system and identify limits up to which style-transfer can help preserve anonymity.

We use three variants of our model in the following study. The first model uses the semantic encoding loss described in \secref{sec:semembloss} and is referred to as \emph{FBsem}. The second uses the reconstruction likelihood loss discussed in ~\secref{sec:cycsemloss} instead, and is denoted by \emph{CycML}. Finally, \emph{CycML+Lang} uses both cyclic maximum likelihood and the language smoothing loss described in \secref{sec:langloss}.

\subsection{Quantitative Evaluation}
\label{sec:quantresults}
\begin{table}[t]
\small
\centering
\caption{F1-scores of the attribute classifiers. All of them do well and better than the document-level random chance (0.62 for speech), (0.53 for age), and (0.50 for gender).\vspace{-3mm}}
\label{tab:DiscriminatorPerf}
\hspace*{-5mm}\begin{tabular}{lrrrr}
\toprule
\multirow{2}{*}{\bf Setting}& \multicolumn{2}{c}{Training Set}& \multicolumn{2}{c}{Validation Set}\\\cmidrule(lr){2-3}\cmidrule(lr){4-5}
                          & Sentence& Document& Sentence& Document\\\cmidrule(lr){1-1}\cmidrule(lr){2-3}\cmidrule(lr){4-5}
Speechdata         & 0.84 & 1.00 & 0.68 & 1.00 \\
Blog-age     & 0.76 & 0.92 & 0.74 & 0.88 \\
Blog-gender  & 0.64 & 0.93 & 0.52 & 0.75 \\
 \bottomrule
\end{tabular}\vspace{-1mm}
\end{table}

Before analyzing the performance of our \ant network, we evaluate the attribute classifiers on the three settings we use. For this, we train the attribute classifier model in \secref{sec:auhtclassifier} on all three settings.
\Tableref{tab:DiscriminatorPerf} shows the F1-scores of the attribute classifiers on the training and the validation splits of the blog and the speech datasets. Document-level scores are obtained from accumulating the class log-probability scores on each sentence in a document before picking the maximum scoring class as the output label. We also tried hard voting to accumulate sentence level decisions, and observed that the hard voting results follow the same trend across datasets and splits.%

On the smaller political speech dataset, the attribute classifier is able to easily discriminate between the two authors, Barack Obama and Donald Trump, achieving perfect F1-score of $1.0$ on both the training and the validation splits. The model also performs well on the age-group classification, achieving F1-score of 0.88 on the validation set at the document-level. 
Gender classification turns out to be the hardest to generalize, with a significant drop in F1-score on the validation set compared to the training set (down to 0.75 from 0.93).
In all three tasks, the performance on sentence-level is worse than on document-level classification.
Document-level classification also generalizes better with less difference between training and validation set F1-scores in~\Tableref{tab:DiscriminatorPerf}.
Henceforth, we will use document-level F1-score as our primary metric when evaluating the effectiveness of \ant networks.

\begin{table*}[t]
\small
\centering
\caption{Performance of the style transfer anonymization in fooling the classifiers, across the three settings. F1 (lower is better) and Meteor (higher is better). F1-scores below chance levels are shown in italics. 
}
\label{tab:TransfPerfdatasets}
\begin{tabular}{lrrrrrrrrr}
\toprule
\multirow{2}{*}{\bf Model}& \multicolumn{3}{c}{Blog-age data }& \multicolumn{3}{c}{Blog-gender data}& \multicolumn{3}{c}{Speech dataset}\\\cmidrule(lr){2-4}\cmidrule(lr){5-7}\cmidrule(lr){8-10}
                          & Sent F1& Doc F1 &Meteor & Sent F1& Doc F1 &Meteor& Sent F1& Doc F1&Meteor\\\cmidrule(lr){1-1}\cmidrule(lr){2-4}\cmidrule(lr){5-7}\cmidrule(lr){8-10}
Random classifier            & 0.54   & 0.54   & - &  0.53 & 0.5 & - & 0.60 & 0.61  & -\\\cmidrule(lr){1-1}\cmidrule(lr){2-4}\cmidrule(lr){5-7}\cmidrule(lr){8-10}
Original text                & 0.74   & 0.88   & 1.0 &  0.68 & 0.75 & 1.0 & 0.68 & 1.00 & 1.0\\
Autoencoder         & 0.69   & 0.85   & 0.77 &  0.49 &  0.61 & 0.78  & \em0.26 & \emph{0.12} &  0.32\\\cmidrule(lr){1-1}\cmidrule(lr){2-4}\cmidrule(lr){5-7}\cmidrule(lr){8-10}
\ant FBsem                     & \em0.43   &\emph{0.08}    & 0.40 &  \em0.45 &  \emph{0.39} & 0.53& \em0.12 & \emph{0.00} & 0.28\\
\ant CycML                     & \em0.49   & \emph{0.20}   & 0.57 &  \em0.41&  \emph{0.44}&\bf 0.79 &  \em0.11 & \emph{0.00} & \bf0.29 \\
\ant CycML+Lang                & \em0.54   & \emph{0.32}   &\bf 0.69 & \em0.44& \emph{0.39}&\bf 0.79 & \em0.12 & \emph{0.00} &\bf 0.29\\
 \bottomrule
\end{tabular}\vspace{-1mm}
\end{table*}
\subsubsection{Quantifying Privacy Gains from \ant}
We evaluate the performance of our \ant network using the attribute classifiers presented in the previous section. 
To obfuscate the authorship of a document, the \ant network is run on every sentence in a document to transfer style from the true attribute to the target one. For example, to mask age of a teenager written document we use the \ant network $Z_{Teen-Adult}$. Style-transferred document is input to the attribute classifier of the source attribute and F1-score of this classifier is computed. This measures the privacy effectiveness of the style transfer. Meteor score is computed between the source sentence and the \ant output, to measure the semantic similarity.%

\Tableref{tab:TransfPerfdatasets} shows these results in the three settings. On the small speech dataset all methods, including the autoencoder baseline described in~\secref{sec:base}, successfully fool the attribute classifier. They all obtain F1-scores below the chance-level, 
with our \ant networks doing better. However the meteor scores of all models is significantly lower than in the blog dataset, indicating significant amount of semantic loss in the process of anonymization.

On the larger blog dataset, the autoencoder baseline fails to fool the attribute classifier, with only a small drop in F1-score of $0.03$ (from $0.88$ to $0.85$) in case of age and $0.14$ in case of gender (from $0.75$ to $0.61$) %
Our \ant models however do much better, with all of them being able to drop the F1-score below the random chance. 

The \emph{FBsem} model using semantic encoder loss achieves the largest privacy gain, by decreasing the F1-scores from $0.88$ to $0.08$ in case of age and from $0.75$ to $0.39$ in case of gender. This model however suffers from poor meteor scores, indicating the sentences produced after the style transfer are no longer similar to the input.

The model using reconstruction likelihood to enforce semantic consistency, \emph{CycML}, fares much better in meteor metric in both age and gender style transfer. It is still able to fool the classifier, albeit with smaller drops in F1-scores (still below random chance). Finally, with addition of the language smoothing loss (\emph{CycML+Lang}), we see a further improvement in meteor in the blog-age setting, while the performance remains similar to \emph{CycML} on blog-gender setting and the speech dataset. 
However, the language smoothing model \emph{CycML+Lang} fares better in human evaluation discussed in \secref{sec:humaneval} and also produces better qualitative samples as will be seen in \secref{sec:qualresults}.

\myparagraph{Generalization to other classifiers:} \ntext{An important question to answer if \ant is to be applied to protect the privacy of author attributes, is how well it performs against unseen NLP based adversaries ? 
To test this we trained ten different attribute classifiers networks on the blog-age setting. These networks vary in architectures (LSTM, CNN and LSTM+CNN) and hyper-parameters (number of layers and number of units), but all of them achieve good performance in predicting the age attribute. The networks were chosen to reflect real-world architecture choices used for text classification. Results from evaluating the text generated by the \ant networks using these ``holdout'' classifiers are shown in~\Tableref{tab:TransfPerfholdout}. The column ``mean'' shows the mean performance of the ten classifiers and ``max'' shows the score of best performing classifier}

\ntext{Holdout classifiers have good performance on the original text, achieving mean $0.85$ document-level F1-score. \Tableref{tab:TransfPerfholdout} shows that all three \ant networks generalize well and are able to drop the document F1-score of the holdout classifiers to the random chance level (0.54 for age dataset). They perform slightly worse than on the seen LSTM classifier, but are able to significantly drop the performance of all the holdout classifiers (mean F1 score drops from 0.85 to 0.53 or below). This is a strong empirical evidence that the transformations applied by the \ant networks are not specific to the classifier they are trained with, but can also generalize to other adversaries.}

We conclude that the proposed \ant networks are able to fool the attribute classifiers on all three tested tasks and also show generalization ability to fool classifier architectures not seen during training.

\myparagraph{Different operating points :} Our \ant model offers the ability to obtain multiple different style-transfer outputs by simply sampling from the models distribution. This is useful as different text samples might have different levels of semantic similarity and privacy effectiveness. Having multiple samples allows users to choose the level of semantic similarity vs privacy trade-off they prefer.

\begin{figure}
\centering
    \includegraphics[width=0.8\columnwidth]{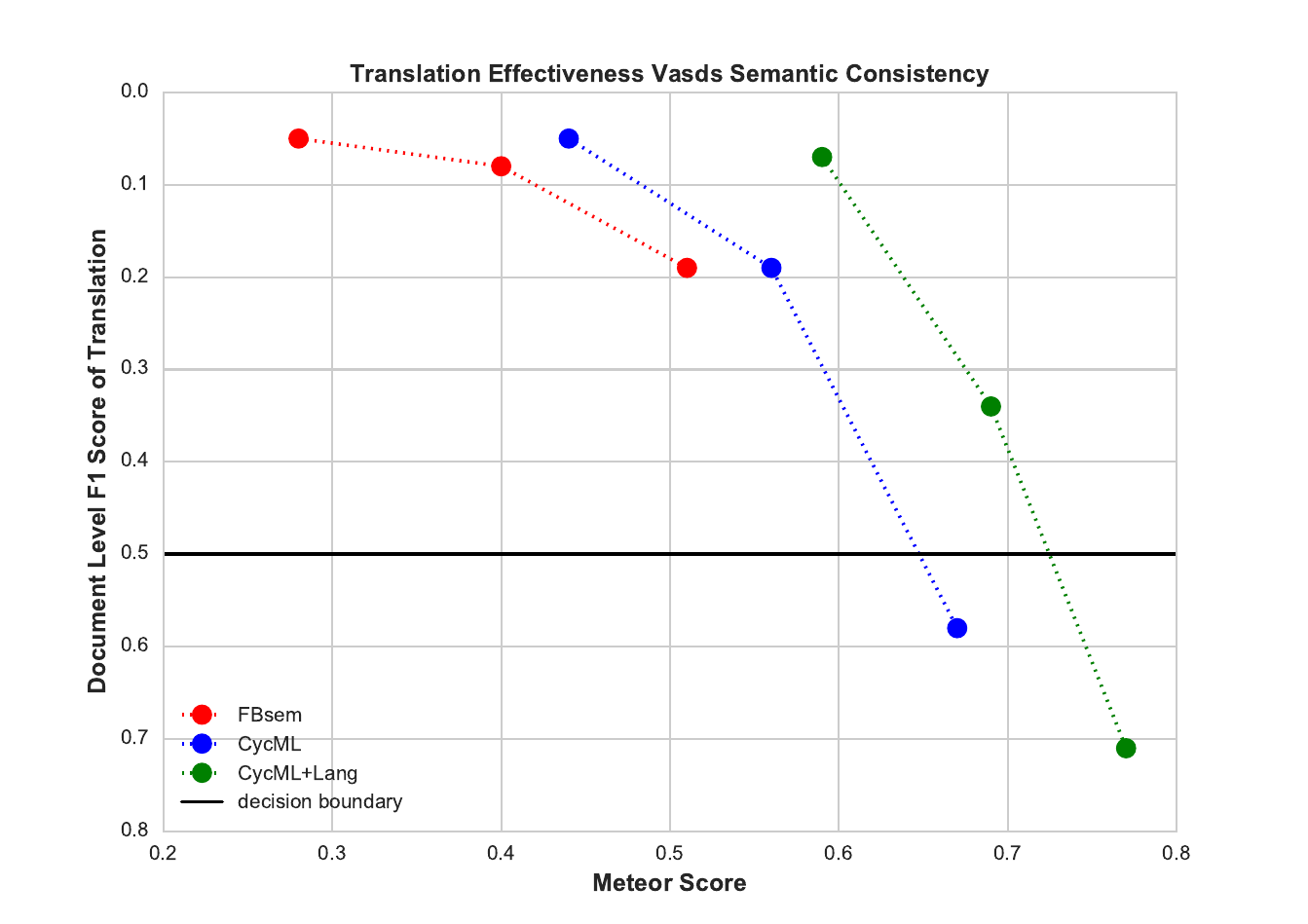}\vspace{-1mm}
    \caption{Operating points of \ant models on test set.\vspace{-4mm}}
\label{fig:OperatingPoints}
\end{figure}

We illustrate this in \Figref{fig:OperatingPoints}. Here five samples are obtained from each \ant model for each sentence in the test set. By choosing the sentence with minimum, maximum or random 
meteor scores, we can obtain a trade-off between semantic similarity and privacy. We see that while the \emph{FBsem} model offers limited variability, \emph{CycML+LangLoss} offers a wide range of choices of operating points. All operating points of \emph{CycML+LangLoss} achieve better meteor score than 0.5, which indicates this model preserves the semantic similarity well.

\begin{table}[t]
\small
\centering
\caption{\ant anonymization fooling unseen classifiers, on blogdata (age). Columns are doc-level F1 score.\vspace*{-5mm}} 
\label{tab:TransfPerfholdout}
\hspace*{-5mm}\begin{tabular}{lrrr}
\toprule
\multirow{2}{*}{\bf Model}& \multirow{2}{*}{\shortstack[c]{Seen \\Classifier}}& \multicolumn{2}{c}{Holdout Classifiers}\\\cmidrule(lr){3-4}
                          &   & Mean & Max\\\cmidrule(lr){1-1}\cmidrule(lr){2-2}\cmidrule(lr){3-4}
Original text           & 0.88 &  0.85& 0.87 \\
Autoencoder             & 0.85 & 0.83 & 0.84 \\\cmidrule(lr){1-1}\cmidrule(lr){2-2}\cmidrule(lr){3-4}
\ant FBsem              & \em0.08 & \em0.19 & \em0.31\\
\ant CycML              & \em0.20 & \em0.41& 0.58\\
\ant CycML+Lang         & \em0.32 & \em0.53 & 0.62\\
 \bottomrule
\end{tabular}\vspace{-1mm}
\end{table}

\subsubsection{Human Judgments for Semantic Consistency}
\label{sec:humaneval}
In machine translation and image captioning literature, it is well known that automatic semantic similarity evaluation metrics like meteor are only reliable to a certain extent. Evaluation from human judges is still the gold-standard with which models can be reliably compared.

Accordingly, we conduct human evaluations to judge the semantic similarity preserved by our \ant networks. The evaluations were conducted on a subset of 745 random sentences from the test split of the blog-age dataset. First, output from different \ant models is obtained for the 745 test sentences. If any model generates identical sentences to the input, this model is ranked first automatically without human evaluation. Note that, in some cases, multiple models can achieve rank-1, when they all produce identical outputs. The cases without any identical sentences to the input are evaluated using human annotators on Amazon Mechanical Turk~(AMT). An annotator is shown one input sentence and multiple style-transfer outputs and is asked to pick the output sentence which is closest in meaning to the input sentence. Three unique annotators are shown each test sample and majority voting is used to determine the model which ranks first. Cases with no majority from human evaluators are excluded.

The main goal of the study is to identify which of the three \ant networks performs best in terms of semantic similarity according to human judges. We also compare the best of our three systems to the baseline model based on Google machine translation, discussed in \secref{sec:base}.

For the machine translation baseline, we obtain style-transferred texts from four different language round-trips. We started with English$\,\to\,$German$\,\to\,$French$\,\to\,$English, and obtained three more versions with incrementally adding Spanish, Finnish and finally Armenian languages into the chain before the translation back to English. 

\begin{figure*}
\CenterFloatBoxes
\begin{floatrow}
\ffigbox[0.34\columnwidth][]{%
    \includegraphics[width=1.0\columnwidth]{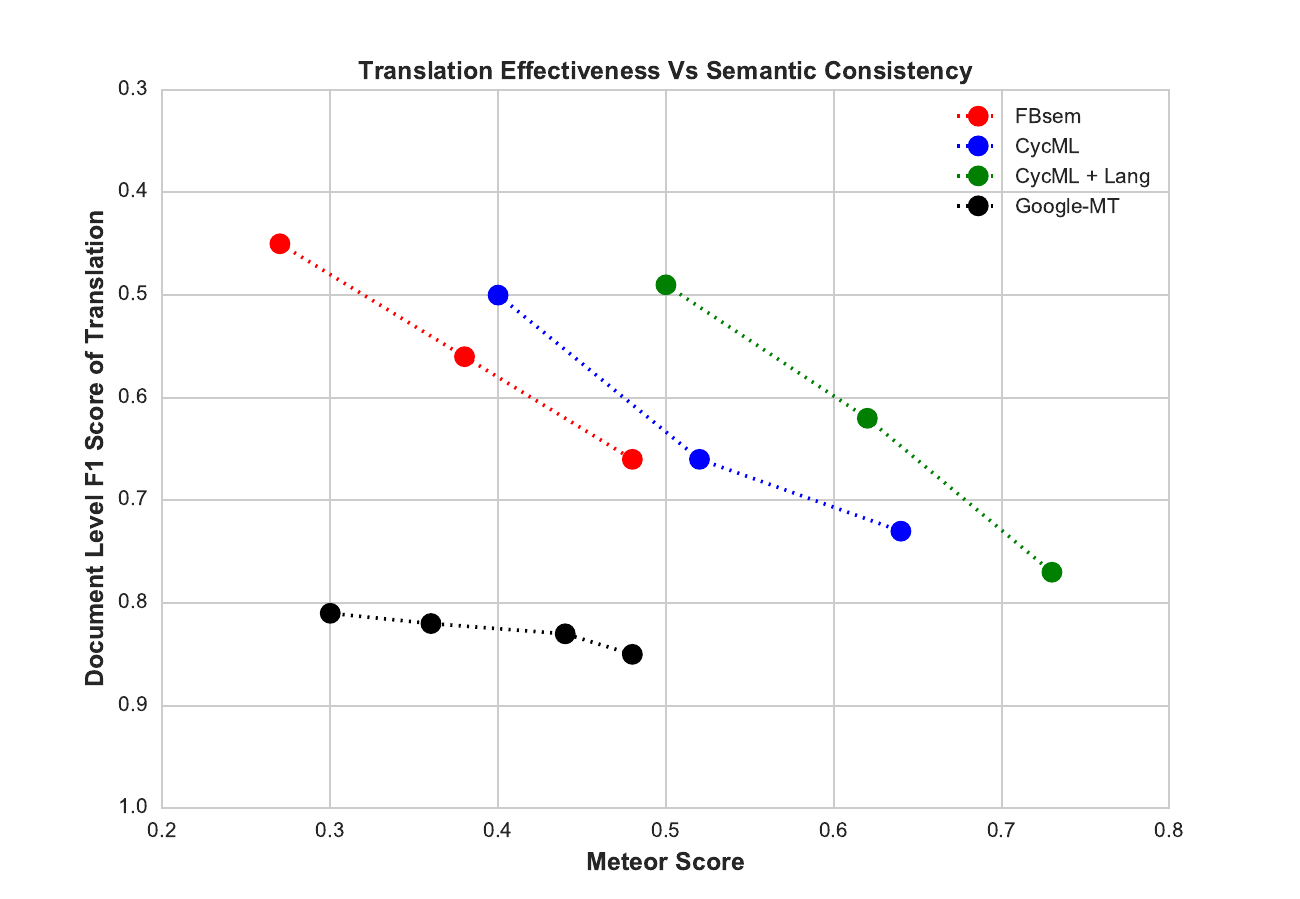}}
    {\caption{Privacy and semantic consistency of \ant and the Google MT baseline on the human eval test set\vspace{-3mm}}\label{fig:humanEffect}}%
\killfloatstyle
\vbox to 2.6cm{%
\footnotesize
\ffigbox[0.66\columnwidth][]{%
\begin{tabular}{lrrr}
\toprule
\textbf{Operating Point} & \bf FBsem&\bf CycML &\bf CycML + Lang \\ \cmidrule(lr){1-1}\cmidrule(lr){2-4}
  0.66           & 32.02& 39.75 &\hfill \bf 57.87   \\
  0.5        & 15.03& 31.68 &\hfill \bf 50.74   \\
 \bottomrule
\end{tabular}\vspace{-2mm}}{\captionof{table}{Human evaluation to judge semantic similarity. Three variants of our model are compared. Numbers show the \% times the model ranked first. Can add to more than 100\% as multiple models can have rank-1.\\\vspace{-2mm}}\label{tab:humanOur}}
\ffigbox[0.66\columnwidth][]{%
\begin{tabular}{lcc}
\toprule
\bf Comparison  &\bf \ant CycML + Lang &\bf GoogleMT\\ \cmidrule(lr){1-1}\cmidrule(lr){2-3}
Operating point & 0.79 & 0.85\\\cmidrule(lr){1-1}\cmidrule(lr){2-3}%
\% Rank 1 & \bf 59.46 & 43.76\\
 \bottomrule
\end{tabular}\vspace{-1mm}
}{\captionof{table}{Human evaluation of our best model and the Google MT baseline.\vspace{-3mm}}\label{tab:humanGoogle}}
}
\end{floatrow}
\end{figure*}

To pick the operating points for the human evaluation study, we compare the performance of these four machine translation baselines and our three models on the human-evaluation test set in \Figref{fig:humanEffect}. Note that here we show sentence-level F1 score on the y-axis as the human-evaluation test set is too small for document-level evaluation. We see that none of the Google machine translation baselines are able to fool the attribute classifiers. The model with 5-hop translation achieves best (lowest) F1-score of $0.81$ which is only slightly less than the input data F1-score of $0.9$. This model also achieves significantly worse meteor score than any of our \ant models. 

We conduct human evaluation for our style-transfer models on two operating points of $0.5$ F1-score and $0.66$ F1-scores, to obtain human judgments at two different levels of privacy effectiveness as shown in \tableref{tab:humanOur}. We see that the model \emph{CycML+Lang} outperforms the other two models at both operating points. \emph{CycML+Lang} wins $50.74\%$ of the time ~(ignoring ties) at operating point $0.5$ and $57.87\%$ of the time at operating point $0.66$. These results combined with quantitative evaluation discussed in \secref{sec:quantresults} confirm that the cyclic ML loss combined with the language model loss gives the best trade-off between semantic similarity and privacy effectiveness.

Finally, we conduct human evaluation between the \emph{CycML+Lang} model operating at $0.79$ and the Google machine translation baseline with 3 hops. The operating point is chosen so that the two models are closest to each other in privacy effectiveness and meteor score.
Results in \tableref{tab:humanGoogle} show that our model wins over the GoogleMT baseline by approximately $16\%$ ($59.46\%$ vs $43.76\%$ rank1) on semantic similarity as per human judges, while still having better privacy effectiveness. This is largely because our \ant model learns not to change the input text if it is already ambiguous for the attribute classifier, and only makes changes when necessary. In contrast, changes made by GoogleMT round trip are not optimized towards maximizing privacy gain, and can change the input text even when no change is needed.

\subsection{Qualitative Analysis}%
\label{sec:qualresults}

In this section we analyze some qualitative examples of anonymized text produced by our \ant model and try to identify strengths and weaknesses of this approach. Then we analyze the performance of the \ant network on different levels of input difficulty. We use the attribute classifiers' score as a proxy measure of the input text difficulty. If the text is confidently correctly classified (with classification score of~$1.0$) by the attribute classifier, then the \ant network has to make significant changes to fool the classifier. If it is already misclassified, the style-transfer network should ideally not make any changes.

\subsubsection{Examples of Style Transfer for anonymization}
\begin{table*}[th]
\begin{center}
\centering
\setlength{\tabcolsep}{2pt}
\footnotesize
\renewcommand{\arraystretch}{1.3}
\begin{tabular}{x{0.01\columnwidth}x{0.43\columnwidth}x{0.04\columnwidth}x{0.46\columnwidth}x{0.04\columnwidth}}
\toprule
\# & \textbf{Input: Teen}&A(x) &\textbf{Output: Adult} &A(x) \\\cmidrule(lr){1-1}\cmidrule(lr){2-3}\cmidrule(lr){4-5}
 1     & and \ul{yeh}... it's raining lots now                           & 0.97 & and \ul{ooh}... it's raining lots now                                      & 0.23 \\
 1     & \ul{yeahh...} i never let anyone really know how i'm feeling.  & 0.94 & \ul{anyhow,} i never let anyone really know how i'm feeling .        & 0.24 \\
 1     &  \ul{yeh}, it's just goin ok here too! & 0.95 &  \ul{alas}, it's just goin ok here too! &  0.30 \\
 1     & \ul{would} i go so far to say that i love her?                  & 0.52 & \ul{will} i go so far to say that i love her?                   & 0.36\\
 2     &\ul{wad} a nice day.. spend almost the whole afternoon doing work!  & 0.99 &\ul{definitely} a nice day.. spend almost the whole afternoon doing work! & 0.19 \\
 2     & \ul{wadeva} told u secrets \ul{wad} did u do ?                & 0.98 & \ul{perhaps} told u secrets \ul{why} did u do ?                    & 0.49 \\ 
 2     & i don't know \ul{y} i even went into \ul{dis} relationship      & 0.92 & i don't know \ul{why} i even went into \ul{another} relationship .    & 0.33 \\
 2     & i have \ul{nuthing} else to say about this \ul{horrid} day.     & 0.79 &  i have \ul{ofcourse} else to say about this \ul{accountable} day.              & 0.08 \\
 3     & after \ul{school} i \ul{got} my hair cut so it looks nice again.& 1.0  & after \ul{all} i \ul{have} my hair cut so it looks nice again.  & 0.42 \\
 3     & i had an interesting day at \ul{skool}. & 0.97  & i had an interesting day at \ul{wedding}.  & 0.05 \\
\cmidrule(lr){1-1}\cmidrule(lr){2-3}\cmidrule(lr){4-5} 
\#  & \textbf{Input: Adult}                                             & A(x) &\textbf{Output: Teen}                                         & A(x) \\  \cmidrule(lr){1-1}\cmidrule(lr){2-3}\cmidrule(lr){4-5}
 1  & \ul{funnily} \ul{enough} , i do n't care all that much.           & 0.58 & \ul{haha} \ul{besides} , i do n't care all that much.             & 0.05\\
 1  & i \ul{may} go to san francisco state, or i may go back.           & 0.54 & i \ul{shall} go to san francisco state, or i may go back.         & 0.09\\
 1  & i wonder if they 'll \ul{work} out... hard to say.                & 0.52 & i wonder if they 'll \ul{go} out... hard to say.                  & 0.39\\
 2  & one is to mix my exercise order a bit more.                       & 0.97 & one is to mix my \ul{diz} exercise order a bit more.              & 0.08\\   
 2  & ok, \ul{think} i really will go to bed now.                       & 0.79 & ok, \ul{relized} i really will go to bed now.                     & 0.08\\
 3  & my first day going out to see \ul{clients} after vacation.             & 0.98 & my first day going out to see \ul{some1} after vacation.               & 0.04\\
 3  & i'd tell my \ul{wife} how much i love her every time i saw her.   & 0.96 & i'd tell my \ul{crush} how much i love her every time i saw her.  & 0.06 \\   
 3  &  i \ul{do} \ul{believe} all you need is love.                     & 0.58 & i \ul{dont} \ul{think} all you need is love . & 0.11\\
  \bottomrule
\end{tabular}\vspace{-3mm}
\caption{Qualitative examples of anonymization through style transfer in the blog-age setting. Style transfer in both direction is shown along with the attribute classifier score of the source attribute.\vspace{-3mm}}
\label{tab:qualexamples}
\end{center}
\end{table*}
\begin{table}[th]
\begin{center}
\centering
\setlength{\tabcolsep}{2pt}
\footnotesize
\renewcommand{\arraystretch}{1.3}
\hspace*{-3mm}\begin{tabular}{x{0.52\columnwidth}x{0.53\columnwidth}}
\toprule
\textbf{Input: Obama} & \textbf{Output: Trump} \\\cmidrule(lr){1-1}\cmidrule(lr){2-2}
we \ul{can} do this because we are MISC. & we \ul{will} do that because we are MISC.\\ 
we \ul{can} do better than that. & we \ul{will} do that better than \ul{anybody}.\\
it's not about \ul{reverend} PERSON.& it's not about \ul{crooked} PERSON.\\
but i'm going to \ul{need} your \ul{help}. & but i'm going to \ul{fight} \ul{for} your \ul{country}.\\
so that's my \ul{vision}. &so that's my \ul{opinion}.\\
their \ul{situation} is getting worse. & their \ul{media} is getting worse.\\
i'm \ul{kind} of the \ul{term} PERSON because \ul{i} \ul{do} care. & i'm \ul{tired} of the \ul{system} of PERSON \ul{PERSON} because \ul{they} \ul{don't} care.\\
that's what \ul{we} \ul{need} to change. & that's what \ul{she} \ul{wanted} to change.\\
that's how our \ul{democracy works}. & that's how our \ul{horrible horrible} \ul{trade deals}.\\
  \bottomrule
\end{tabular}\vspace{-3mm}
\caption{Qualitative examples of style transfer on the speech dataset from Obama to Trump's style\vspace{-3mm}}
\label{tab:qualexamplesSpeech}
\end{center}
\end{table}
\Tableref{tab:qualexamples} shows the results of our \ant model \emph{CycML+Lang} applied to some example sentences in the blog-age setting. Style transfer in both directions, teenager to adult and adult to teenager, is shown along with the corresponding source attribute classifier scores. The examples illustrate some of the common changes made by the model and are grouped into three categories for analysis (\# column in \tableref{tab:qualexamples}). 

\myparagraph{\# 1. Using synonyms:}
The \ant network often uses synonyms to change the style to target attribute. This is seen in style transfers in both directions, teen to adult and adult to teen in category \# 1 samples in \tableref{tab:qualexamples}. We can see the model replacing ``yeh'' with ``ooh'',  ``would'' with ``will'', ``...'' with ``,'' and so on when going from teen to adult, and replacing ``funnily enough'' with ``haha besides'', ``work out'' with ``go out'' and so on when changing from adult to teen.
We can also see that the changes are not static, but depend on the context. For example ``yeh'' is replaced with ``alas'' in one instance and with ``ooh'' in another. These changes do not alter the meaning of the sentence too much, but fool the attribute classifiers thereby providing privacy to the author attribute.%

\myparagraph{\# 2. Replacing slang words:}
When changing from teen to adult, \ant often replaces the slang words or incorrectly spelled words with standard English words, as seen in category \#2 in \tableref{tab:qualexamples}. For example, replacing ``wad'' (what) with ``definitely'', ``wadeva'' with ``perhaps'' and ``nuthing'' with ``ofcourse''. The opposite effect is seen when going from adult to teenager, with addition of ``diz'' (this) and replacing  of ``think'' with ``relized'' (realized). These changes are learned entirely from the data, and would be very hard to encode explicitly in a rule-based system due to the variety in slangs and spelling mistakes.

\myparagraph{\# 3. Semantic changes:}
One failure mode of \ant is when the input sentence has semantic content which is significantly more biased to the author's class. These examples are shown in category \#3 in \tableref{tab:qualexamples}. For example, when an adult author mentions his ``wife'', the \ant network replaces it with ``crush'', altering the meaning of the input sentence. Some common entity pairs where this behavior is seen are with (\emph{school}$\leftrightarrow$\emph{work}), (\emph{class}$\leftrightarrow$\emph{office}), (\emph{dad}$\leftrightarrow$\emph{husband}), (\emph{mum}$\leftrightarrow$\emph{wife}), %
and so on. Arguably, in such cases, there is no obvious solution to mask the identity of the author without altering these obviously biased content words.

On the smaller speech dataset however, the changes made by the \ant model alter the semantics of the sentences in some cases. Few example style transfers from Obama to Trump's style are shown in \Tableref{tab:qualexamplesSpeech}. 
We see that \ant inserts hyperbole (``better than anybody'', ``horrible horrible'', ``crooked''), references to ``media'' and ``system'', all salient features of Trump's style. We see that the style-transfer here is quite successful, sufficient to completely fool the identity classifier as was seen in \tableref{tab:TransfPerfdatasets}.  However, and somewhat expectedly, the semantics of the input sentence is generally lost. A possible cause is that the attribute classifier is too strong on this data, owing to the small dataset size and the highly distinctive styles of the two authors, and to fool them the \ant network learns to make drastic changes to the input text.
\subsubsection{Performance Across Input Difficulty} 
\begin{figure}[h]
\CenterFloatBoxes
\begin{floatrow}
\hspace{-2mm}
\ffigbox[0.5\columnwidth][]{%
\includegraphics[width=1.2\columnwidth]{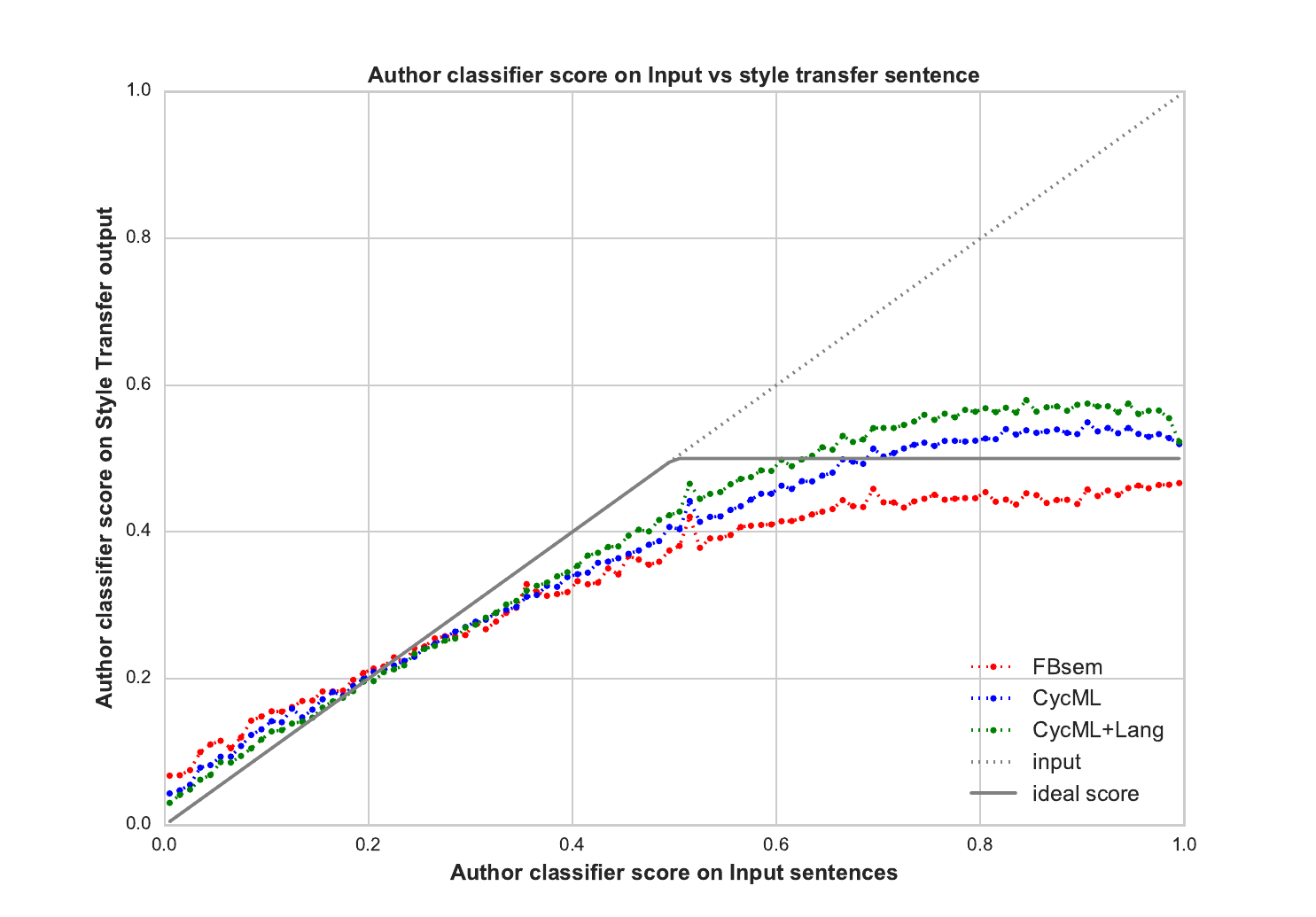}}
    {\caption{Output Privacy vs Privacy on Input.\vspace{-4mm}}\label{fig:TargetClassScore}}%
\hspace{-3mm}
\ffigbox[0.5\columnwidth][]{%
\includegraphics[width=1.2\columnwidth]{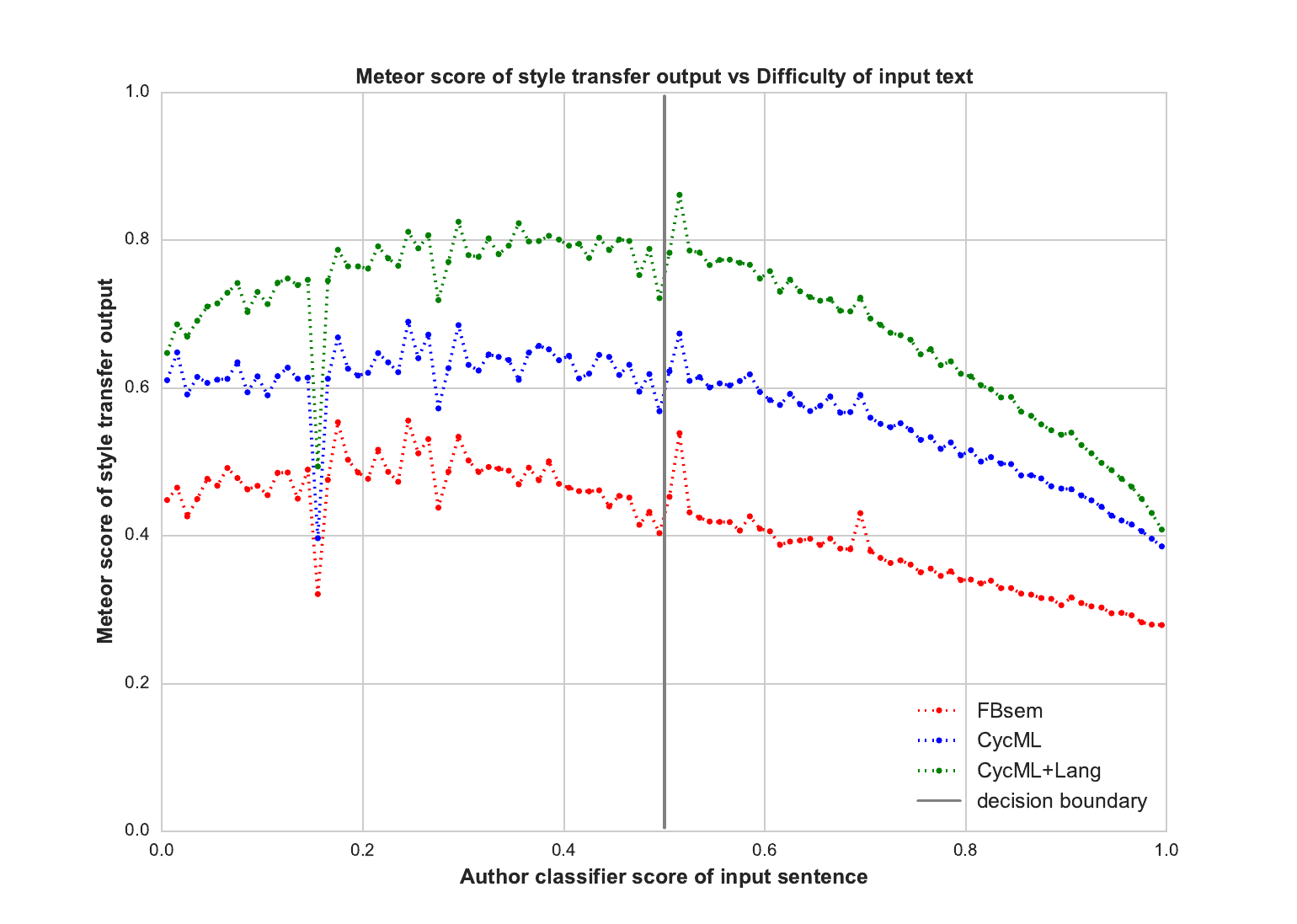}}
    {\caption{Meteor score plotted against input difficulty.\vspace{-4mm}}\label{fig:InputDiffivsMeteorScore}}%
\end{floatrow}
\end{figure}

\figref{fig:TargetClassScore} compares the attribute classifier score on the input sentence and the \ant output. Ideally we want all the\ant outputs to score below the decision boundary, while also not increasing the classifier score compared to input text. This ``ideal score'' is shown as grey solid line. We see that for the most part all three \ant models are below or close to this ideal line. As the input text gets more difficult (increasing attribute classifier score), the \emph{CycML} and \emph{CycML+Lang} slightly cross  above the ideal line, but still provide significant improvement over the input text (drop in classifier score of about $\sim 0.45$).
\begin{figure}[h]
\includegraphics[width=0.8\columnwidth]{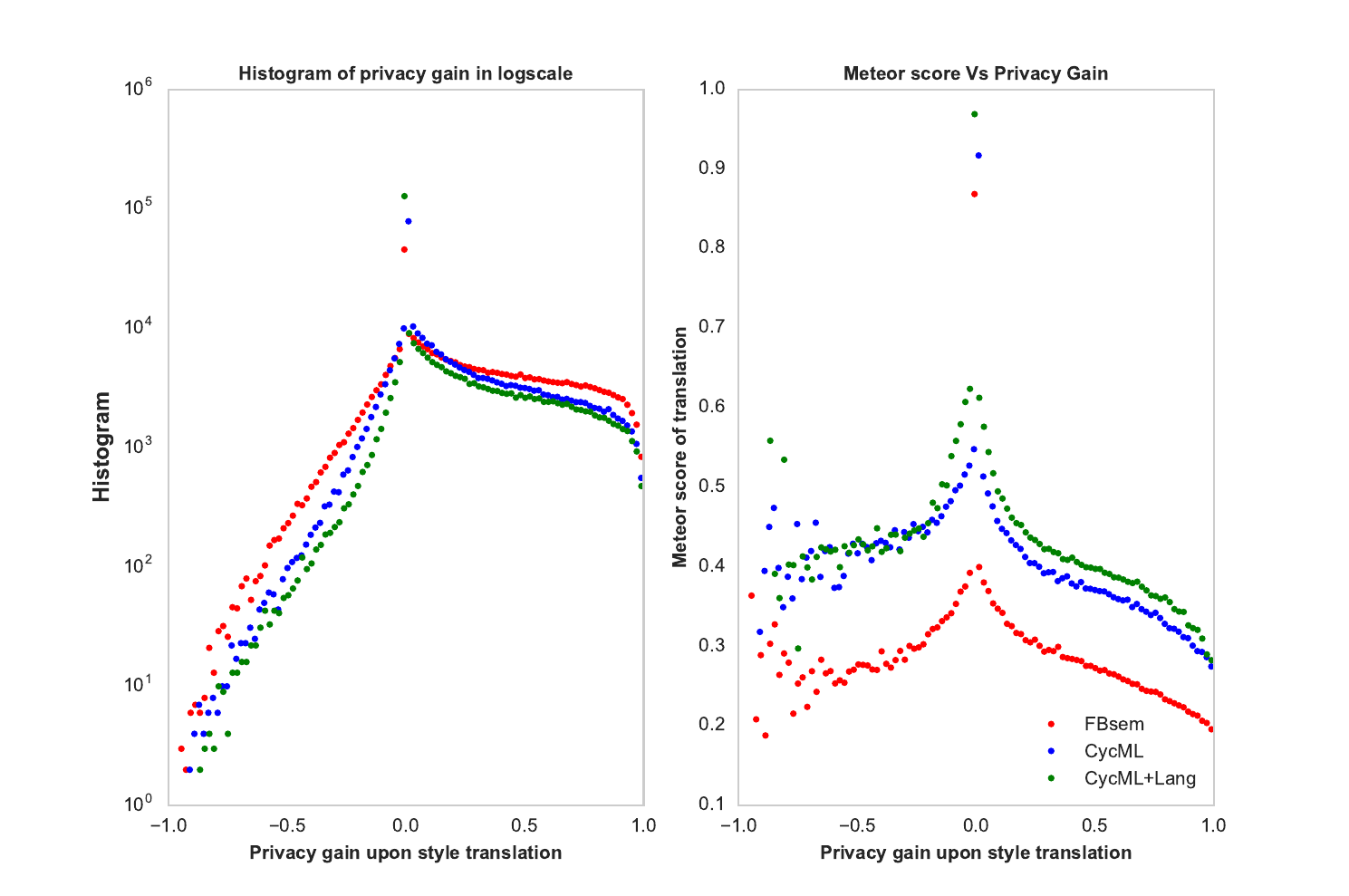}
\caption{Histogram of privacy gain (left side) is shown alongside comparison of meteor score vs privacy gains.\vspace{-4mm}}\label{fig:PrivacyGainvsMeteorScore}
\end{figure}

Now, we analyze how much of input semantics is preserved with increasing difficulty. \Figref{fig:InputDiffivsMeteorScore} plots the meteor score of the \ant output against the difficulty of input text. We see that the meteor is high for sentences already across the decision boundary. These are easy cases, where the \ant networks need not intervene. As the input gets more difficult, the meteor score of the \ant output drops, as the network needs to do more changes to be able to fool the attribute classifier. The \emph{CycML+Lang} model fares better than the other two models, with consistently higher meteor across the difficulty spectrum.

\Figref{fig:PrivacyGainvsMeteorScore} shows the histogram of privacy gain across the test set. Privacy gain is the difference between the attribute classifier score on the input and the \ant network output. We see that majority of transformations by the \ant networks leads to positive privacy gains, with only a small fraction leading to negative privacy gains. This is promising given that this histogram is over all the 500k sentences in the test set. Meteor score plotted against privacy gain shown in \figref{fig:PrivacyGainvsMeteorScore}, again confirms that large privacy gains comes with a trade-off of loss in semantics.

\section{Conclusions}
We presented a novel fully automatic method for protecting privacy sensitive attributes of an author against NLP based attackers. Our solution, the \ant network, is developed using a novel application of adversarial training to machine translation networks to learn to protect private attributes. The \ant network achieves this by learning to perform style-transfer without paired data. 

\ant offers a new data driven approach to authorship obfuscation. The flexibility of this end-to-end trainable model means it can adapt to new attack methods and datasets.
Experiments on three different attributes namely age, gender and identity, showed that the \ant network is able to effectively fool the attribute classifiers in all the three settings. We also show that the \ant network also performs well against multiple unseen classifier architectures. This strong empirical evidence suggests that the method is likely to be effective %
against previously unknown NLP adversaries.%

We developed a novel solution to preserve the meaning of input text using likelihood of reconstruction. Semantic similarity (quantified by meteor score) of the \ant network remains high for easier sentences, which do not contain obvious give-away words (school, work, husband etc.), but is lower on difficult sentences indicating the network effectively learns to identify and apply the right magnitude of change.
The \ant network can be operated at different points on the privacy-effectiveness and semantic-similarity trade-off curve, and thus offers flexibility to the user.
The experiments on the political speech data show the limits to which style transfer based approach can be used to hide attributes. On this challenging data with very distinct styles by the two authors, our method effectively fools the identity classifier but achieves this by altering the semantics of the input text.

\ntext{ In future work we would like to explore generator architectures to extend \ant framework to structured data like code, to protect against code stylometric attacks.}

\section*{Acknowledgment}
This research was supported in part by the German Research Foundation (DFG CRC 1223). We would also like to thank Yang Zhang, Ben Stock and Sven Bugiel for helpful feedback.
\FloatBarrier
\bibliographystyle{IEEEtran}
\bibliography{biblioLong,paper}

\begin{thebibliography}{10}
\providecommand{\url}[1]{#1}
\csname url@samestyle\endcsname
\providecommand{\newblock}{\relax}
\providecommand{\bibinfo}[2]{#2}
\providecommand{\BIBentrySTDinterwordspacing}{\spaceskip=0pt\relax}
\providecommand{\BIBentryALTinterwordstretchfactor}{4}
\providecommand{\BIBentryALTinterwordspacing}{\spaceskip=\fontdimen2\font plus
\BIBentryALTinterwordstretchfactor\fontdimen3\font minus
  \fontdimen4\font\relax}
\providecommand{\BIBforeignlanguage}[2]{{%
\expandafter\ifx\csname l@#1\endcsname\relax
\typeout{** WARNING: IEEEtran.bst: No hyphenation pattern has been}%
\typeout{** loaded for the language `#1'. Using the pattern for}%
\typeout{** the default language instead.}%
\else
\language=\csname l@#1\endcsname
\fi
#2}}
\providecommand{\BIBdecl}{\relax}
\BIBdecl

\bibitem{juola2008authorship}
P.~Juola \emph{et~al.}, ``Authorship attribution,'' \emph{Foundations and
  Trends{\textregistered} in Information Retrieval}, vol.~1, no.~3, pp.
  233--334, 2008.

\bibitem{stamatatos2009survey}
E.~Stamatatos, ``A survey of modern authorship attribution methods,''
  \emph{Journal of the Association for Information Science and Technology},
  vol.~60, no.~3, pp. 538--556, 2009.

\bibitem{ruder2016character}
S.~Ruder, P.~Ghaffari, and J.~G. Breslin, ``Character-level and multi-channel
  convolutional neural networks for large-scale authorship attribution,''
  \emph{arXiv preprint arXiv:1609.06686}, 2016.

\bibitem{argamon2009automatically}
S.~Argamon, M.~Koppel, J.~W. Pennebaker, and J.~Schler, ``Automatically
  profiling the author of an anonymous text,'' \emph{Communications of the
  ACM}, vol.~52, no.~2, pp. 119--123, 2009.

\bibitem{overdorf2016blogs}
R.~Overdorf and R.~Greenstadt, ``Blogs, twitter feeds, and reddit comments:
  Cross-domain authorship attribution,'' \emph{Proceedings on Privacy Enhancing
  Technologies}, vol. 2016, no.~3, pp. 155--171, 2016.

\bibitem{narayanan2012feasibility}
A.~Narayanan, H.~Paskov, N.~Z. Gong, J.~Bethencourt, E.~Stefanov, E.~C.~R.
  Shin, and D.~Song, ``On the feasibility of internet-scale author
  identification,'' in \emph{Security and Privacy (SP), 2012 IEEE Symposium
  on}.\hskip 1em plus 0.5em minus 0.4em\relax IEEE, 2012, pp. 300--314.

\bibitem{jkrowling}
\BIBentryALTinterwordspacing
P.~Juola. (2013) How a computer program helped show {J.K.} rowling write a
  cuckoo’s calling. [Online]. Available: \url{https://goo.gl/mkZai1}
\BIBentrySTDinterwordspacing

\bibitem{morgan2017predicting}
A.~A. Morgan-Lopez, A.~E. Kim, R.~F. Chew, and P.~Ruddle, ``Predicting age
  groups of twitter users based on language and metadata features,'' \emph{PloS
  one}, vol.~12, no.~8, p. e0183537, 2017.

\bibitem{Ikeda:2013}
K.~Ikeda, G.~Hattori, C.~Ono, H.~Asoh, and T.~Higashino, ``Twitter user
  profiling based on text and community mining for market analysis,''
  \emph{Know.-Based Syst.}, vol.~51, 2013.

\bibitem{makazhanov2014predicting}
A.~Makazhanov, D.~Rafiei, and M.~Waqar, ``Predicting political preference of
  twitter users,'' \emph{Social Network Analysis and Mining}, vol.~4, no.~1, p.
  193, 2014.

\bibitem{likesHelpedTrump}
\BIBentryALTinterwordspacing
H.~Grassegger and M.~Krogerus. (2017) The data that turned the world upside
  down. [Online]. Available:
  \url{https://motherboard.vice.com/en_us/article/mg9vvn/how-our-likes-helped-trump-win}
\BIBentrySTDinterwordspacing

\bibitem{brennan2012adversarial}
M.~Brennan, S.~Afroz, and R.~Greenstadt, ``Adversarial stylometry:
  Circumventing authorship recognition to preserve privacy and anonymity,''
  \emph{ACM Transactions on Information and System Security (TISSEC)}, vol.~15,
  no.~3, p.~12, 2012.

\bibitem{afroz2012detecting}
S.~Afroz, M.~Brennan, and R.~Greenstadt, ``Detecting hoaxes, frauds, and
  deception in writing style online,'' in \emph{Security and Privacy (SP), 2012
  IEEE Symposium on}.\hskip 1em plus 0.5em minus 0.4em\relax IEEE, 2012, pp.
  461--475.

\bibitem{McDonaldACSG12}
A.~W. McDonald, S.~Afroz, A.~Caliskan, A.~Stolerman, and R.~Greenstadt, ``Use
  fewer instances of the letter" i": Toward writing style anonymization.'' in
  \emph{Privacy Enhancing Technologies}, vol. 7384.\hskip 1em plus 0.5em minus
  0.4em\relax Springer, 2012, pp. 299--318.

\bibitem{castro:2017}
D.~Castro, R.~Ortega, and R.~Mu{\~n}oz, ``{Author Masking by Sentence
  Transformation---Notebook for PAN at CLEF 2017},'' in \emph{{CLEF 2017
  Evaluation Labs and Workshop -- Working Notes Papers}}, Sep. 2017.

\bibitem{keswani2016author}
Y.~Keswani, H.~Trivedi, P.~Mehta, and P.~Majumder, ``Author masking through
  translation.'' in \emph{CLEF (Working Notes)}, 2016, pp. 890--894.

\bibitem{CaliskanTranslate2012}
A.~Caliskan and R.~Greenstadt, ``Translate once, translate twice, translate
  thrice and attribute: Identifying authors and machine translation tools in
  translated text,'' in \emph{2012 IEEE Sixth International Conference on
  Semantic Computing}, Sept 2012, pp. 121--125.

\bibitem{goodfellow2014generative}
I.~Goodfellow, J.~Pouget-Abadie, M.~Mirza, B.~Xu, D.~Warde-Farley, S.~Ozair,
  A.~Courville, and Y.~Bengio, ``Generative adversarial nets,'' in
  \emph{Advances in Neural Information Processing Systems (NIPS)}, 2014, pp.
  2672--2680.

\bibitem{seqToseq2014}
I.~Sutskever, O.~Vinyals, and Q.~V. Le, ``Sequence to sequence learning with
  neural networks,'' in \emph{Advances in neural information processing
  systems}, 2014, pp. 3104--3112.

\bibitem{xu2012paraphrasing}
W.~Xu, A.~Ritter, B.~Dolan, R.~Grishman, and C.~Cherry, ``Paraphrasing for
  style,'' \emph{Proceedings of COLING 2012}, pp. 2899--2914, 2012.

\bibitem{afroz2014doppelganger}
S.~Afroz, A.~C. Islam, A.~Stolerman, R.~Greenstadt, and D.~McCoy,
  ``Doppelg{\"a}nger finder: Taking stylometry to the underground,'' in
  \emph{Security and Privacy (SP), 2014 IEEE Symposium on}.\hskip 1em plus
  0.5em minus 0.4em\relax IEEE, 2014, pp. 212--226.

\bibitem{caliskan2015anonymizing}
A.~Caliskan-Islam, R.~Harang, A.~Liu, A.~Narayanan, C.~Voss, F.~Yamaguchi, and
  R.~Greenstadt, ``De-anonymizing programmers via code stylometry,'' in
  \emph{USENIX Security Symposium}, 2015.

\bibitem{abbasi2008writeprints}
A.~Abbasi and H.~Chen, ``Writeprints: A stylometric approach to identity-level
  identification and similarity detection in cyberspace,'' \emph{ACM
  Transactions on Information Systems (TOIS)}, vol.~26, no.~2, p.~7, 2008.

\bibitem{bagnall2015author}
D.~Bagnall, ``Author identification using multi-headed recurrent neural
  networks,'' \emph{arXiv preprint arXiv:1506.04891}, 2015.

\bibitem{obfGary2006}
G.~Kacmarcik and M.~Gamon, ``Obfuscating document stylometry to preserve author
  anonymity,'' in \emph{Proceedings of the COLING/ACL on Main conference poster
  sessions}.\hskip 1em plus 0.5em minus 0.4em\relax Association for
  Computational Linguistics, 2006, pp. 444--451.

\bibitem{karadzhov2017case}
G.~Karadzhov, T.~Mihaylova, Y.~Kiprov, G.~Georgiev, I.~Koychev, and P.~Nakov,
  ``The case for being average: A mediocrity approach to style masking and
  author obfuscation,'' in \emph{International Conference of the Cross-Language
  Evaluation Forum for European Languages}.\hskip 1em plus 0.5em minus
  0.4em\relax Springer, 2017, pp. 173--185.

\bibitem{bahdanau2014neural}
D.~Bahdanau, K.~Cho, and Y.~Bengio, ``Neural machine translation by jointly
  learning to align and translate,'' \emph{Proceedings of the International
  Conference on Learning Representations (ICLR)}, 2014.

\bibitem{wu2016google}
Y.~Wu, M.~Schuster, Z.~Chen, Q.~V. Le, M.~Norouzi, W.~Macherey, M.~Krikun,
  Y.~Cao, Q.~Gao, K.~Macherey \emph{et~al.}, ``Google's neural machine
  translation system: Bridging the gap between human and machine translation,''
  \emph{arXiv preprint arXiv:1609.08144}, 2016.

\bibitem{shen2017style}
T.~Shen, T.~Lei, R.~Barzilay, and T.~Jaakkola, ``Style transfer from
  non-parallel text by cross-alignment,'' \emph{To appear in NIPS}, 2017.

\bibitem{hochreiter1997long}
S.~Hochreiter and J.~Schmidhuber, ``Long short-term memory,'' \emph{Neural
  computation}, vol.~9, no.~8, 1997.

\bibitem{mikolov13nips}
T.~Mikolov, I.~Sutskever, K.~Chen, G.~S. Corrado, and J.~Dean, ``Distributed
  representations of words and phrases and their compositionality,'' in
  \emph{Advances in Neural Information Processing Systems (NIPS)}, 2013.

\bibitem{pennington2014glove}
J.~Pennington, R.~Socher, and C.~Manning, ``Glove: Global vectors for word
  representation,'' in \emph{Proceedings of the 2014 conference on empirical
  methods in natural language processing (EMNLP)}, 2014, pp. 1532--1543.

\bibitem{weiss2017sequence}
R.~J. Weiss, J.~Chorowski, N.~Jaitly, Y.~Wu, and Z.~Chen,
  ``Sequence-to-sequence models can directly transcribe foreign speech,''
  \emph{arXiv preprint arXiv:1703.08581}, 2017.

\bibitem{ma2016end}
X.~Ma and E.~Hovy, ``End-to-end sequence labeling via bi-directional
  {LSTM-CNNs-CRF},'' in \emph{Proceedings of the Annual Meeting of the
  Association for Computational Linguistics (ACL)}, 2016, pp. 1064--1074.

\bibitem{Shetty2017iccv}
R.~Shetty, M.~Rohrbach, L.~A. Hendricks, M.~Fritz, and B.~Schiele, ``Speaking
  the same language: Matching machine to human captions by adversarial
  training,'' in \emph{Proceedings of the IEEE International Conference on
  Computer Vision (ICCV)}, 2017.

\bibitem{jang2016categorical}
E.~Jang, S.~Gu, and B.~Poole, ``Categorical reparameterization with
  gumbel-softmax,'' \emph{Proceedings of the International Conference on
  Learning Representations (ICLR)}, 2016.

\bibitem{CycleGAN2017}
J.-Y. Zhu, T.~Park, P.~Isola, and A.~A. Efros, ``Unpaired image-to-image
  translation using cycle-consistent adversarial networks,'' \emph{Proceedings
  of the IEEE International Conference on Computer Vision (ICCV)}, 2017.

\bibitem{conneau2017supervised}
A.~Conneau, D.~Kiela, H.~Schwenk, L.~Barrault, and A.~Bordes, ``Supervised
  learning of universal sentence representations from natural language
  inference data,'' in \emph{Proceedings of the Conference on Empirical Methods
  in Natural Language Processing (EMNLP)}, 2017.

\bibitem{kiros2015skip}
R.~Kiros, Y.~Zhu, R.~R. Salakhutdinov, R.~Zemel, R.~Urtasun, A.~Torralba, and
  S.~Fidler, ``Skip-thought vectors,'' in \emph{Advances in neural information
  processing systems}, 2015, pp. 3294--3302.

\bibitem{bowman2015large}
S.~R. Bowman, G.~Angeli, C.~Potts, and C.~D. Manning, ``A large annotated
  corpus for learning natural language inference,'' \emph{Proceedings of the
  Conference on Empirical Methods in Natural Language Processing (EMNLP)},
  2015.

\bibitem{pytorch}
\BIBentryALTinterwordspacing
Pytorch framework. [Online]. Available: \url{http://pytorch.org/}
\BIBentrySTDinterwordspacing

\bibitem{tieleman2012lecture}
T.~Tieleman and G.~Hinton, ``Lecture 6.5-rmsprop: Divide the gradient by a
  running average of its recent magnitude,'' \emph{COURSERA: Neural networks
  for machine learning}, vol.~4, no.~2, pp. 26--31, 2012.

\bibitem{schler2006effects}
J.~Schler, M.~Koppel, S.~Argamon, and J.~W. Pennebaker, ``Effects of age and
  gender on blogging.'' in \emph{AAAI spring symposium: Computational
  approaches to analyzing weblogs}, vol.~6, 2006, pp. 199--205.

\bibitem{woolley2008american}
\BIBentryALTinterwordspacing
J.~T. Woolley and G.~Peters. (1999) The american presidency project. [Online].
  Available: \url{http://www.presidency.ucsb.edu}
\BIBentrySTDinterwordspacing

\bibitem{finkel2005incorporating}
J.~R. Finkel, T.~Grenager, and C.~Manning, ``Incorporating non-local
  information into information extraction systems by gibbs sampling,'' in
  \emph{Proceedings of the Annual Meeting of the Association for Computational
  Linguistics (ACL)}, 2005, pp. 363--370.

\bibitem{denkowski-lavie:2014:Meteor}
M.~Denkowski and A.~Lavie, ``Meteor universal: Language specific translation
  evaluation for any target language,'' in \emph{Proceedings of the Ninth
  Workshop on Statistical Machine Translation}.\hskip 1em plus 0.5em minus
  0.4em\relax ACL, 2014, pp. 376--380.

\bibitem{maddison2016concrete}
C.~J. Maddison, A.~Mnih, and Y.~W. Teh, ``The concrete distribution: A
  continuous relaxation of discrete random variables,'' \emph{Proceedings of
  the International Conference on Learning Representations (ICLR)}, 2016.

\end{thebibliography}
\vfill\eject
\begin{appendices}
\section{Appendix - Differentiability of discrete samples}
\label{sec:appendixA}

To obtain an output sentence sample $\tilde{s_y}$ from the \ant network $Z_{xy}$, we can sample from the distribution $p(\tilde{w}_t|s_x)$, shown in \eqref{eqn:genOut}, repeatedly until a special `END' token is sampled. This naive sampling though is not suitable for training $Z$ within a GAN framework as sampling from multinomial distribution, $p(\tilde{w}_t|s_x)$, is not differentiable. 

To make sampling differentiable we follow the approach used in~\cite{Shetty2017iccv} and use the Gumbel-Softmax approximation~\cite{jang2016categorical} to obtain differentiable soft samples from $p(\tilde{w}_t|s_x)$. The gumbel-softmax approximation includes two parts. First, the re-parametrization trick using the gumbel random variable is applied to make the process of sampling from a multinomial distribution differentiable with respect to the probabilities $p(\tilde{w}_t|s_x)$. Next, softmax is used to approximate the arg-max operator to obtain ``soft'' samples instead of one-hot vectors. This makes the samples themselves differentiable. Thus, the gumbel-softmax approximation allows differentiating through sentence samples from the \ant network enabling end-to-end GAN training. Further details on gumbel-softmax approximation can be found in~\cite{jang2016categorical,maddison2016concrete}.
\end{appendices}

\end{document}